\documentclass[pdftex,twocolumn,3]{jour3}          

\usepackage[utf8]{inputenc}
\usepackage{epsf}
\usepackage{latexsym,amssymb,euscript}
\usepackage{widetext}
\usepackage[dvips]{graphicx}
\usepackage[numbers,sort&compress]{natbib}
\usepackage{amsmath}
\usepackage{nicefrac}
\usepackage{slashed}
\usepackage{booktabs}
\usepackage{hyperref}
\usepackage{braket}
\usepackage{chngcntr}
\usepackage{bm}
\usepackage{bbold}
\usepackage{graphics}
\usepackage{graphicx}
\usepackage{mciteplus}
\usepackage{bbold}
\usepackage{pdfpages}
\usepackage[titletoc]{appendix}
\graphicspath{{./figures/}}
\hypersetup{
 linktocpage = true,
 urlcolor = urlblue,
 colorlinks = true,
 linkcolor = urlblue,
 anchorcolor = urlblue,
 citecolor = urlblue,
 pdfstartview = {XYZ null null 1.25} 
           }
\usepackage[left=2cm, right=2cm]{geometry}
\usepackage{pstricks}
\usepackage{color}
\usepackage{xcolor}
\definecolor{urlblue}{rgb}{0.2,0.4,0.7}
\definecolor{citegreen}{rgb}{0,0.4,0.2}
\definecolor{linkred}{rgb}{0.9,0.2,0.1}
\usepackage{float}
\usepackage{academicons}
\definecolor{orcidlogocol}{HTML}{A6CE39}
\usepackage{fancyhdr}
\newcommand{\drv}{{\rm d}}

\newcommand{\MSb}{\overline{\rm MS}}

\newcommand{\CnNLA}{{\cal C}_n^{\rm NLA}}

\newcommand{\DY}{\Delta Y}
\newcommand{\JPsi}{J/\psi}
\newcommand{\Yps}{\Upsilon}
\newcommand{\Q}{{\cal Q}}

\newcommand{\tcite}[1]{~\cite{#1}}
\newcommand{\tref}[1]{~\ref{#1}}
\newcommand{\eref}[1]{~\eqref{#1}}

\newcommand{\tarr}{
\begin{array}}
\newcommand{\earr}{\end{array}}

\newcommand{\orcidFGC}{\href{https://orcid.org/0000-0003-3299-2203}{\includegraphics[scale=0.1]{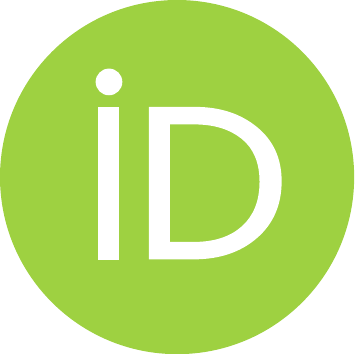}}}

\newcommand{\orcidMF}{\href{https://orcid.org/0000-0002-2408-2210}{\includegraphics[scale=0.1]{figures/logo-orcid.pdf}}}

\smartqed  

\journalname{}

\begin{document}

\title{Inclusive $J/\psi$ and $\Upsilon$ emissions from single-parton fragmentation in hybrid high-energy and collinear factorization
}

\subtitle{}

\author{
Francesco Giovanni Celiberto
\thanksref{e1,addr1,addr2,addr3} \orcidFGC
\and
Michael Fucilla
\thanksref{e2,addr4,addr5,addr6} \orcidMF
}

\thankstext{e1}{{\it e-mail}:
\href{mailto:fceliberto@ectstar.eu}{fceliberto@ectstar.eu}}
\thankstext{e2}{{\it e-mail}:
\href{mailto:michael.fucilla@unical.it }{michael.fucilla@unical.it} (corresponding author)}

\institute{European Centre for Theoretical Studies in Nuclear Physics and Related Areas (ECT*), I-38123 Villazzano, Trento, Italy\label{addr1}
\and
Fondazione Bruno Kessler (FBK),
I-38123 Povo, Trento, Italy\label{addr2}
\and
INFN-TIFPA Trento Institute of Fundamental Physics and Applications,
I-38123 Povo, Trento, Italy\label{addr3}
\and
Dipartimento di Fisica, Universit\`a della Calabria, I-87036 Arcavacata di Rende, Cosenza, Italy\label{addr4}
\and
Istituto Nazionale di Fisica Nucleare, Gruppo collegato di Cosenza, I-87036 Arcavacata di Rende, Cosenza, Italy\label{addr5}
\and
Université Paris-Saclay, CNRS, IJCLab, 91405 Orsay, France\label{addr6}
\vspace{0.75cm}
}

\date{\today}

\maketitle


\section*{Abstract}

We present a novel study on the inclusive production of a heavy quarkonium ($J/\psi$ or $\Upsilon$), in association with a light-flavored jet, as a test field of the high-energy QCD dynamics. The large transverse momenta at which the two final-state objects are detected permits us to perform an analysis in the spirit of the variable-flavor number scheme (VFNS), in which the cross section for the hadroproduction of a light parton is convoluted with a perturbative fragmentation function that describes the transition from a light quark to a heavy hadron.
The quarkonium collinear fragmentation function is built as a product between a short-distance coefficient function, which encodes the resummation of DGLAP type logarithms, and a non-perturbative long-distance matrix element (LDME), calculated in the non-relativistic QCD (NRQCD) fra\-me\-work. Our theoretical setup is the hybrid high-ener\-gy and col\-li\-ne\-ar factorization, where the standard col\-li\-ne\-ar approach is supplemented by the resummation of leading and next-to-leading energy-type logarithms \emph{\`a la} BFKL. We propose this reaction as a suitable channel to probe the production mechanisms of quarkonia at high energies and large transverse momenta and to possibly unveil the transition region from the heavy-quark pair production mechanism to the single-parton fragmentation one.
\vspace{0.50cm} \hrule
\vspace{0.50cm}

{
 \setlength{\parindent}{0pt}
 \textsc{Keywords}: \vspace{0.15cm} \\ QCD phenomenology \\ high-energy resummation \\ quarkonium production \\  NRQCD fragmentation
\vspace{0.50cm} \hrule
\vspace{0.50cm}
}



\section{Introduction}
In this work we focus on the high-energy inclusive hadroproduction of a heavy vector mesons. The energies reachable at the Large Hadron Collider (LHC) and at new-generation hadron accelerators\tcite{Chapon:2020heu,Anchordoqui:2021ghd,Feng:2022inv,Celiberto:2022rfj,Hentschinski:2022xnd,Accardi:2012qut,AbdulKhalek:2021gbh,Khalek:2022bzd,Acosta:2022ejc,AlexanderAryshev:2022pkx,Celiberto:2016yek,Celiberto:2016jse,Brunner:2022usy,Arbuzov:2020cqg,Abazov:2021hku,Bernardi:2022hny,Amoroso:2022eow,Celiberto:2018hdy,Klein:2020nvu,2064676,MuonCollider:2022xlm,Aime:2022flm} permits us to access kinematic regions where $\sqrt{s} \gg m_{\cal Q}$, with $m_{\cal Q}$ the vector mass. In this case, the so-called \emph{Regge-Gribov} or \emph{semi-hard} regime\tcite{Gribov:1983ivg,Celiberto:2017ius} of QCD, characterized by $\sqrt{s} \gg \{Q \} \gg \Lambda_{\rm QCD}$, is reached.\footnote{Here $s$ is the center-of-mass energy squared, $\{Q \}$ represents the (set of) perturbative scale(s) characterizing the process, and $\Lambda_{\rm QCD}$ stands for the QCD scale parameter.} 
As it is well known, in this regime, large energy-type logarithms appear to all orders in perturbation theory with a power increasing with the $\alpha_s$ order, thus spoiling the convergence of the perturbative series obtained within the standard fixed-order approach. A resummation to all orders, that takes into account the effect of these logarithms, it is strictly necessary to build concrete predictions in this regime. The most powerful formalism for the description of the semi-hard QCD sector is the Balitsky--Fadin--Kuraev--Lipatov (BFKL) approach~\cite{Fadin:1975cb,Kuraev:1976ge,Kuraev:1977fs,Balitsky:1978ic}, which is valid in the leading approximation (LLA), \emph{i.e.} $\alpha_s^n \ln (s/Q^2)^n$ terms are resummed, and within the next-to-leading approximation (NLA), \emph{i.e.} also $\alpha_s^{n+1} \ln (s/Q^2)^n$ terms are also included.

BFKL-resummed cross sections are elegantly portrayed by a high-energy convolution between a universal, energy-dependent Green’s function, and two impact factors depicting the transition from each incoming particle to the outgoing object(s) produced in its fragmentation region. The BFKL Green's function satisfies an integral evolution equation, whose kernel is known up to NLO for any fixed, not growing with $s$, momentum transfer $t$ and for any possible two-gluon color state in the $t$-channel\tcite{Fadin:1998py,Ciafaloni:1998gs,Fadin:1998jv,Fadin:2000kx,Fadin:2000hu,Fadin:2004zq,Fadin:2005zj}. Impact factors instead depend on processes, so they represent the most challenging part of the calculation. Few of them are known within NLO accuracy.

Combining pairwise the available NLO impact factors, a number of semi-hard (inclusive)
reactions can be described within the BFKL approach in the NLA and predictions can be
formulated, mainly in the form of azimuthal correlations or transverse momentum distributions,
most of them accessible to current experiments at the LHC. These reactions include: the inclusive detection of two light jets featuring large transverse momenta and well separated in rapidity (Mueller--Navelet channel\tcite{Mueller:1986ey}), for which several phenomenological studies have appeared so far\tcite{Colferai:2010wu,Caporale:2012ih,Ducloue:2013hia,Ducloue:2013bva,Caporale:2013uva,Caporale:2014gpa,Colferai:2015zfa,Caporale:2015uva,Ducloue:2015jba,Celiberto:2015yba,Celiberto:2015mpa,Celiberto:2016ygs,Celiberto:2016vva,Caporale:2018qnm,deLeon:2021ecb,Celiberto:2022gji}), the inclusive emission of a light di-hadron system\tcite{Celiberto:2016hae,Celiberto:2016zgb,Celiberto:2017ptm,Celiberto:2017uae,Celiberto:2017ydk}, multi-jet tags\tcite{Caporale:2015vya,Caporale:2015int,Caporale:2016soq,Caporale:2016vxt,Caporale:2016xku,Celiberto:2016vhn,Caporale:2016djm,Caporale:2016pqe,Chachamis:2016qct,Chachamis:2016lyi,Caporale:2016lnh,Caporale:2016zkc,Chachamis:2017vfa,Caporale:2017jqj}, hadron-jet\tcite{Bolognino:2018oth,Bolognino:2019cac,Bolognino:2019yqj,Celiberto:2020rxb}, Higgs-jet\tcite{Celiberto:2020tmb,Celiberto:2021fjf,Celiberto:2021tky,Celiberto:2021txb,Celiberto:2021xpm,Celiberto:2022fgx}, Drell--Yan-plus-jet \cite{Golec-Biernat:2018kem}, hadrons with heavy flavor \cite{Boussarie:2017oae,Celiberto:2017nyx,Bolognino:2019ouc,Bolognino:2019yls,Bolognino:2019ccd,Celiberto:2021dzy,Celiberto:2021fdp,Bolognino:2021zco,Bolognino:2022wgl,Celiberto:2022dyf,Celiberto:2022keu,Celiberto:2022zdg}, and heavy-light di-jet systems \cite{Bolognino:2021mrc,Bolognino:2021hxxaux}.

It is well known that phenomenological studies of semi-hard processes feature instabilities of the BFKL series under higher-order corrections and scale variations, preventing us to carry out analyses around \emph{natural} scales of the process, namely the ones dictated by kinematics.
Some scale-optimization procedures, such as the Brodsky--Lepage--Mackenzie (BLM) procedure~\cite{Brodsky:1996sg,Brodsky:1997sd,Brodsky:1998kn,Brodsky:2002ka} one, can help to reduce these instabilities in reactions featuring the emissions of light jets and/or hadrons\tcite{Ducloue:2013bva,Caporale:2014gpa,Celiberto:2016hae,Celiberto:2017ptm}. 
Nevertheless, the optimal-scale values predicted by the procedure are much higher than the natural ones\tcite{Celiberto:2020wpk}. This leads to a lowering of cross sections of one or more orders of magnitude.

A first sign of stability was observed in partial NLA analyses, involving impact factors characterized by the presence of a particle having a large transverse mass, as Higgs bosons~\cite{Celiberto:2020tmb,Celiberto:2022zdg} and heavy-quark jets~\cite{Bolognino:2021mrc}. Until now, however, the lack of Higgs and heavy quark NLO impact factors has prevented the confirmation of these stability in a full NLL analysis.\footnote{The forward Higgs-boson impact factor has been recently calculated at NLO (see Refs.\tcite{Hentschinski:2020tbi,Celiberto:2022fgx} for novel calculations), but its numerical 
implementation is not yet available.} 

Recently, it was pointed out that stabilization effects in full NLA observables emerge in a variable-flavor number-scheme (VFNS)~\tcite{Mele:1990cw,Cacciari:1993mq} treatment of charmed and bottomed hadrons' production~\cite{Celiberto:2021dzy,Celiberto:2021fdp}. In particular, it has been found that peculiar pattern of VFNS fragmentation functions (FFs), depicting the production of heavy hadrons at large transverse momentum, acts as a
fair stabilizer of the high-energy series.

In this article we investigate the high-energy behavior of the inclusive hadroproduction of a $\JPsi$ or a $\Yps$ accompanied by a light-quark jet at the LHC.
Both the meson and the jet feature large transverse momenta, and they are well separated in rapidity. 
The large required transverse momenta makes valid using the variable-flavor number-scheme (VFNS) approach. In this spirit, we will combine three main objects: 1) parton distribution functions (PDFs), 2) cross sections for the production of a light parton, 3) FFs describing the transition from light quarks to heavy bound states ($\JPsi$ or $\Yps$). The latters represent new ingredients enriching our hybrid high-energy and collinear factorization already set as the reference formalism for the description of inclusive two-particle semi-hard emissions.
Quarkonium FFs at the initial scale are constructed, within the framework of NRQCD, as a product between a short-distance coefficient function, depicting the transition from a heavy quark to the Fock state of the produced hadron, and the corresponding LDME.
At this point, the DGLAP evolution comes into play. Indeed, when the transverse momenta of the parton produced in the initial hard scattering is sufficiently larger than the production threshold, the heavy quark can be dynamically generated through a cascade of collinear emissions. Therefore, what within the context of the VFNS we could define as perturbative FF is constructed in two steps: the first in which the transition from light to heavy parton takes place and one in which the heavy parton produces the quarkonium according to the dictates of NRQCD
We will employ a recent NLO determination for heavy-quark to $\JPsi$ and $\Yps$ FFs\tcite{Zheng:2019dfk}.

As already mentioned, the fragmentation mechanism becomes increasingly competitive as well as the transverse momentum of the produced quarkonium grows.
We remind, however, that since we are neglecting the heavy-quark mass at the level of the hard part, our formalism is not suited to properly describe the intermediate region, where $|{\vec p}_T| \sim m_Q$. Here, powers of the $\sqrt{m_Q^2 + |{\vec p}_T|^2}/|{\vec p}_T|$ ratio need to be taken into account. In this sense, our study is complementary to the one proposed in Ref.\tcite{Boussarie:2017oae} (see also preliminary results in Refs.\tcite{Boussarie:2015jar,Boussarie:2016gaq}), where the inclusive semi-hard $\JPsi$-plus-jet production was investigated by making use of the short-distance $(Q \bar Q)$ mechanism for the quarkonium production. 
In such a calculation all powers of the ratio $\sqrt{m_Q^2 + |{\vec p}_T|^2}/|{\vec p}_T|$ are present, but since the final state features no DGLAP evolution, collinear logarithms are not resummed. 

\section{Hybrid high-energy and collinear factorization}
The process under investigation is
\begin{equation}
\label{process}
    p(P_a) + p(P_b) \rightarrow \Q(p_\Q, y_\Q) + X + {\rm jet}(p_J, y_J) \; ,
\end{equation}
where $p(P_{a,b})$ indicate an initial proton with momentum $P_{a,b}$, $\Q(p_\Q, y_\Q)$ stands for $\JPsi$ or $\Yps$ emitted with momentum $p_\Q$ and rapidity $y_\Q$, $p_J$ and $y_J$ are the transverse momenta and rapidity of the produced light jet and $X$ denotes the undetected remnant of the reaction. 
The diffractive semi-hard configuration in the final state is obtained by requiring a large rapidity separation $\DY = y_\Q - y_J$ and high observed transverse momenta, $|\vec p_{\Q,J}|$. 
Moreover, a large transverse-momentum is required in order to ensure of a VFNS tratment of the quarkonium fragmentation.

We introduce the standard Sudakov decomposition for the four-momenta of final-state particles,
\begin{equation}
\begin{split}
& p_{\Q,J} = x_{\Q,J} P_{a,b} + \frac{\vec p_{\Q,J}^{\,2}}{x_{\Q,J} s}P_{b,a} + p_{\Q,J\perp} \; \\ &
{\rm{with}} \hspace{0.5 cm} p_{\Q,J\perp}^2=-\vec p_{\Q,J}^{\,2} \; ,
\end{split}
\label{sudakov}
\end{equation}
and where we choose $P_a$ and $P_b$ as light cone basis, \emph{i.e.} $P_a^2= P_b^2=0$ and $s = 2 (P_a\cdot P_b)$.
The longitudinal momentum fractions, $x_{\Q,J}$, are related to the corresponding rapidities by the relation
\begin{equation}
y_{\Q,J}=\pm\frac{1}{2}\ln\frac{x_{\Q,J}^2 s}
{\vec p_{\Q,J}^2}.    
\end{equation}
In collinear factorization, the LO cross section of our process (Eq.\eref{process}) is given by the convolution of the partonic hard sub-process with the parent-proton PDFs and the quarkonium FFs
\begin{eqnarray}
&& \frac{\drv\sigma^{\rm LO}_{\rm coll.}}{\drv x_\Q\drv x_J\drv ^2\vec p_\Q\drv ^2\vec p_J}
= \sum_{i,j = q,{\bar q},g} \int_0^1 \hspace{-0.20cm} \drv x_i \int_0^1 \hspace{-0.20cm} \drv x_j \\ \nonumber
&\times&  \hspace{-0.10cm} f_i \left(x_a\right) f_j \left(x_b\right)  \hspace{-0.05cm}
\int_{x_\Q}^1 \hspace{-0.15cm} \frac{\drv z}{z}D^{\Q}_{i}\left(\frac{x_\Q}{z}\right) 
\frac{\drv {\hat\sigma}_{i, j}\left(\hat s\right)}
{\drv x_\Q\drv x_J\drv ^2\vec p_\Q\drv ^2\vec p_J}\;.
\label{sigma_collinear}
\end{eqnarray}
Here the $i,j$ indices indicate the parton species (quarks $q = u, d, s, c, b$; antiquarks $\bar q = \bar u, \bar d, \bar s, \bar c, \bar b$; or gluon $g$), $f_{i , j}\left(x_{a,b}, \mu_F \right)$ are the PDFs, while $D^{\Q}_{i}\left(x/ z, \mu_F \right)$ denote the quarkonium FFs; $x_{a,b}$ are the longitudinal momentum fractions of the partons initiating the hard subprocess and $z$ the longitudinal fraction of the single parton that fragments into $\Q$. Finally, $\drv\hat\sigma_{i,j}\left(\hat s \right)$ is the partonic cross section, with $\hat s \equiv x_a x_b s$ the squared center-of-mass energy of the partonic collision.

Contrariwise to the pure collinear treatment,
we build the cross section in the hybrid factorization,
where the high-energy dynamics is genuinely provided by the BFKL approach, and collinear ingredients are then embodied. We decompose the cross section as a Fourier sum of azimuthal-angle
coefficients, $\mathcal{C}_n$, in the following way
\begin{equation}
 \frac{(2\pi)^2 \drv \sigma}{\drv y_\Q \drv y_J \drv \vec p_\Q \drv \vec p_J \drv \phi_\Q \drv \phi_J} \!=\! \left[{\cal C}_0 + 2 \!\sum_{n=1}^\infty\! \cos (n \varphi)\, 
 {\cal C}_n \right] \; ,
 \label{dsigma_Fourier}
\end{equation}
where $\varphi_{\Q,J}$ are azimuthal angles of the detected particles and $\varphi = \phi_\Q - \phi_J - \pi$.
The azimuthal coefficients ${\cal C}_n \equiv \CnNLA$ are computed in the BFKL framework and they contain the resummation of energy logarithms up to the NLA accuracy. In the $\MSb$ renormalization scheme, the NLA expression for $\CnNLA$ reads (see, \emph{e.g.}, Ref.~\cite{Caporale:2012ih})
\begin{equation*}
\begin{split}
 & \CnNLA = \int_0^{2\pi} \drv \phi_\Q \int_0^{2\pi} \drv \phi_J\,
 \cos (n \varphi) \, \\ &
 \times \frac{\drv \sigma_{\rm NLA}}{\drv y_\Q \drv y_J\, \drv |\vec p_\Q| \, \drv |\vec p_J| \drv \phi_\Q \drv \phi_J} = \hspace{-0.10 cm} \int_{-\infty}^{+\infty} \hspace{-0.50 cm} \drv \nu \, \frac{e^{{\DY} \bar \alpha_s(\mu_R) \chi(n,\nu)}}{e^{- \DY}  s}
\end{split}
\end{equation*}
\begin{equation*}
  e^{{\DY} \bar \alpha_s^2(\mu_R)
 \left[\bar\chi(n,\nu)+\frac{\beta_0}{8 N_c}\chi(n,\nu)\left[\frac{10}{3}-\chi(n,\nu)+4\ln\left(\frac{\mu_R}{\sqrt{|\vec p_\Q| |\vec p_J|}}\right)\right]\right]}
\end{equation*}
\begin{equation*}
 \times \, \alpha_s^2(\mu_R) \, 
 \bigg [
 c_\Q^{\rm NLO}(n,\nu,|\vec p_\Q|, x_1)[c_J^{\rm NLO}(n,\nu,|\vec p_J|,x_2)]^*  \end{equation*}
 \begin{equation}
\left.
 + \, \bar \alpha_s^2(\mu_R)
 \, \DY
 \frac{\beta_0}{4 N_c}\chi(n,\nu)f(\nu)
 \right] \; ,
 \label{Cn_NLA_MSb}
\end{equation} 
with $\bar \alpha_s(\mu_R) \equiv \alpha_s(\mu_R) N_c/\pi$, $N_c$ the color number and $\beta_0 = 11N_c/3 - 2 n_f/3$ the first coefficient of the QCD $\beta$-function.
The Lipatov characteristic function is given by
\begin{equation}
\chi\left(n,\nu\right)=2\left\{\psi\left(1\right)-{\rm Re} \left[\psi\left( i\nu+\frac{1}{2}+\frac{n}{2} \right)\right] \right\}
\label{chi}
\end{equation}
where $\psi(z) = \Gamma^\prime(z)/\Gamma(z)$ is the logarithmic derivative of the Gamma function. The $\bar\chi(n,\nu)$ function in Eq.\eref{Cn_NLA_MSb} contains NLO corrections to the BFKL kernel and was calculated in Ref.~\cite{Kotikov:2000pm} (see also Ref.~\cite{Kotikov:2002ab}).
The two functions
\begin{equation}
\begin{split}
c_{\Q,J}^{\rm NLO}(n,\nu,|\vec p\,|,x) &=
c_{\Q,J}(n,\nu,|\vec p\,|,x) \\ & +
\alpha_s(\mu_R) \, \hat c_{\Q,J}(n,\nu,|\vec p\,|,x)
\end{split}
\label{IFs}
\end{equation}
are the impact factors for the production of a heavy-quarkonium state and for the emission of a light jet.
Their LO parts read
\[
c_\Q(n,\nu,|\vec p\,|,x) 
= 2 \sqrt{\frac{C_F}{C_A}}
\frac{(|\vec p\,|^2)^{i\nu}}{|\vec p\,|}\,\int_{x}^1\frac{\drv \zeta}{\zeta}
\left( \frac{\zeta}{x} \right)
^{2 i\nu-1} 
\]
\begin{equation}
\label{LOQIF}
 \times \left[\frac{C_A}{C_F}f_g(\zeta)D_g^\Q\left(\frac{x}{\zeta}\right)
 +\sum_{\alpha=q,\bar q}f_\alpha(\zeta)D_\alpha^\Q\left(\frac{x}{\zeta}\right)\right] 
\end{equation}
and
 \begin{equation}
 \begin{split}
 \label{LOJIF}
 c_J(n,\nu,|\vec p\,|,x) &= 2 \sqrt{\frac{C_F}{C_A}}
 (|\vec p\,|^2)^{i\nu-1/2}\, \\ & \times \left(\frac{C_A}{C_F}f_g(x)
 +\sum_{\beta=q,\bar q}f_\beta(x)\right) \;,
 \end{split}
\end{equation}
respectively. The $f(\nu)$ function is defined in terms of the logarithmic derivative of LO impact factors
\begin{equation}
 f(\nu) = \frac{i}{2} \, \frac{\drv}{\drv \nu} \ln\left(\frac{c_\Q}{c_J^*}\right) + \ln\left(|\vec p_\Q| |\vec p_J|\right) \;.
\label{fnu}
\end{equation}
The remaining functions in Eq.~(\ref{Cn_NLA_MSb}) are the NLO impact-factor corrections, $\hat c_{\Q,J}$.
The NLO correction to the $\Q$ impact factor is calculated in the light-quark limit\tcite{Ivanov:2012iv}. This option is fully consistent with our VFNS treatment, provided that the $p_\Q$ values at work are much larger than the heavy-quark mass.

\section{Quarkonium fragmentation functions}
NLO collinear FF sets for the \emph{direct} $\JPsi$ or $\Yps$ meson production are constructed by taking, as a starting point, a NLO calculation\tcite{Zheng:2019dfk} for the heavy-quark FF portraying the transition $c \to \JPsi$ or the $b \to \Yps$ one, where $c$ ($b$) indistinctly refer to the charm (bottom) quark and its antiquark. 
It basically relies on the NRQCD factorization formalism (see, \emph{e.g.}, Refs.\tcite{Thacker:1990bm,Bodwin:1994jh,QuarkoniumWorkingGroup:2004kpm,Brambilla:2010cs,Pineda:2011dg,Brambilla:2020ojz}), in which the FF function of a parton $i$ fragmenting into a heavy quarkonium $\cal Q$ with longitudinal fraction $z$ is written as
\begin{equation}
 \label{FF_NRQCD}
 D^{\cal Q}_i(z, \mu_F) = \sum_{[n]} {\cal D}^{\cal Q}_{i}(z, \mu_F, [n]) \langle {\cal O}^{\cal Q}([n]) \rangle \;.
\end{equation}
In Eq.\eref{FF_NRQCD}, ${\cal D}_{i}(z, \mu_F, [n])$ is the short-distance perturbative coefficient, while $\langle {\cal O}^{\cal Q}([n]) \rangle$ is the NRQCD LDME. The summation is extended over all quarkonium quantum numbers $[n] \equiv \,^{2S+1}L_J^{(c)}$, in the spectroscopic notation, the $(c)$ superscript identifying the color state, singlet (1) or octet (8).

We consider here only a spin-triplet (vector) and color-singlet quarkonium state, $^3S_1^{(1)}$. The form of the initial-scale FF portraying the constituent heavy-quark to quarkonium transition, $Q \to {\cal Q}$, reads\tcite{Zheng:2019dfk}
\begin{eqnarray}
 D^{\cal Q}_Q(z, \mu_F \equiv \mu_0) 
 &=& D^{\cal Q, {\rm LO}}_Q(z) \\ \nonumber
 &+& \frac{\alpha_s^3(3m_Q)}{m_Q^3} \, |{\cal R}_{\cal Q}(0)|^2 \, \Gamma_Q^{\cal Q, {\rm NLO}}(z) \;,
\label{FF_Q-to-onium}
\end{eqnarray}
with $m_c = 1.5$~GeV or~$m_b = 4.9$~GeV, and the radial wave-function at the origin of the quarkonium state set to\tcite{Eichten:1994gt}
\begin{equation*}
  |{\cal R}_{\JPsi}(0)|^2 = 0.810~{\rm{GeV}}^3
\end{equation*}
or to 
\begin{equation*}
|{\cal R}_{\Yps}(0)|^2 = 6.477~{\rm{GeV}}^3.  
\end{equation*}
The LO initial-scale FF reads\tcite{Braaten:1993mp}
\begin{equation}
 \begin{split}
 D^{\cal Q, {\rm LO}}_Q(z) &= 
 \frac{\alpha_s^2(3m_Q)}{m_Q^3} \, \frac{8z(1-z)^2}{27\pi (2-z)^6} \, |{\cal R}_{\cal Q}(0)|^2 \, \\ & \times (5z^4 - 32z^3 + 72z^2 -32z + 16) \;,    
 \end{split}
 \label{FF_Q-to-onium_LO}
\end{equation}
and the polynomial function $\Gamma_Q^{\cal Q, {\rm NLO}}(z)$ entering the expression for the NLO-FF correction is of the form
\begin{equation}
    \Gamma_Q^{\cal Q, {\rm NLO}}(z) = \sum_{n=0}^{10} c_n z^n \; .
    \label{FF_Gamma_calQ}
\end{equation}
Coefficients of $z$-powers in Eqs.\eref{FF_Gamma_calQ} can be found in Ref.\tcite{Zheng:2019dfk}. They are obtained via a polynomial fit to the numerically-calculated NLO FFs.
Starting from $\mu_F \equiv \mu_0 = 3 m_Q$, in Ref.\tcite{Zheng:2019dfk} a DGLAP-evolved formula for the $D^{\cal Q}_Q(z, \mu_F)$ function was derived and then applied to phenomenological studies of $\JPsi$ and $\Yps$ production via $e^+ e^-$ single inclusive annihilation (SIA). In all predictions we will use this FF, to which, from now, we refer as {\tt ZCW19}.

\section{Phenomenology}
Numerical analysis was carried out by using {\tt JETHAD} modular work package\tcite{Celiberto:2020wpk}.
Sensitivity to scale variation of our predictions is evaluated allowing the ratio $C_\mu = \mu_{R,F}/\mu_N$ to vary from 1/2 to two.
The error related to phase-space multi-dimensional integration is embodied and it is always kept below 1\% by the {\tt JETHAD} integrators.
All calculations of our observables are done in the $\MSb$ scheme. BLM scales are calculated in the MOM scheme.

\begin{figure*}[!t]
\centering
   \includegraphics[scale=0.50,clip]{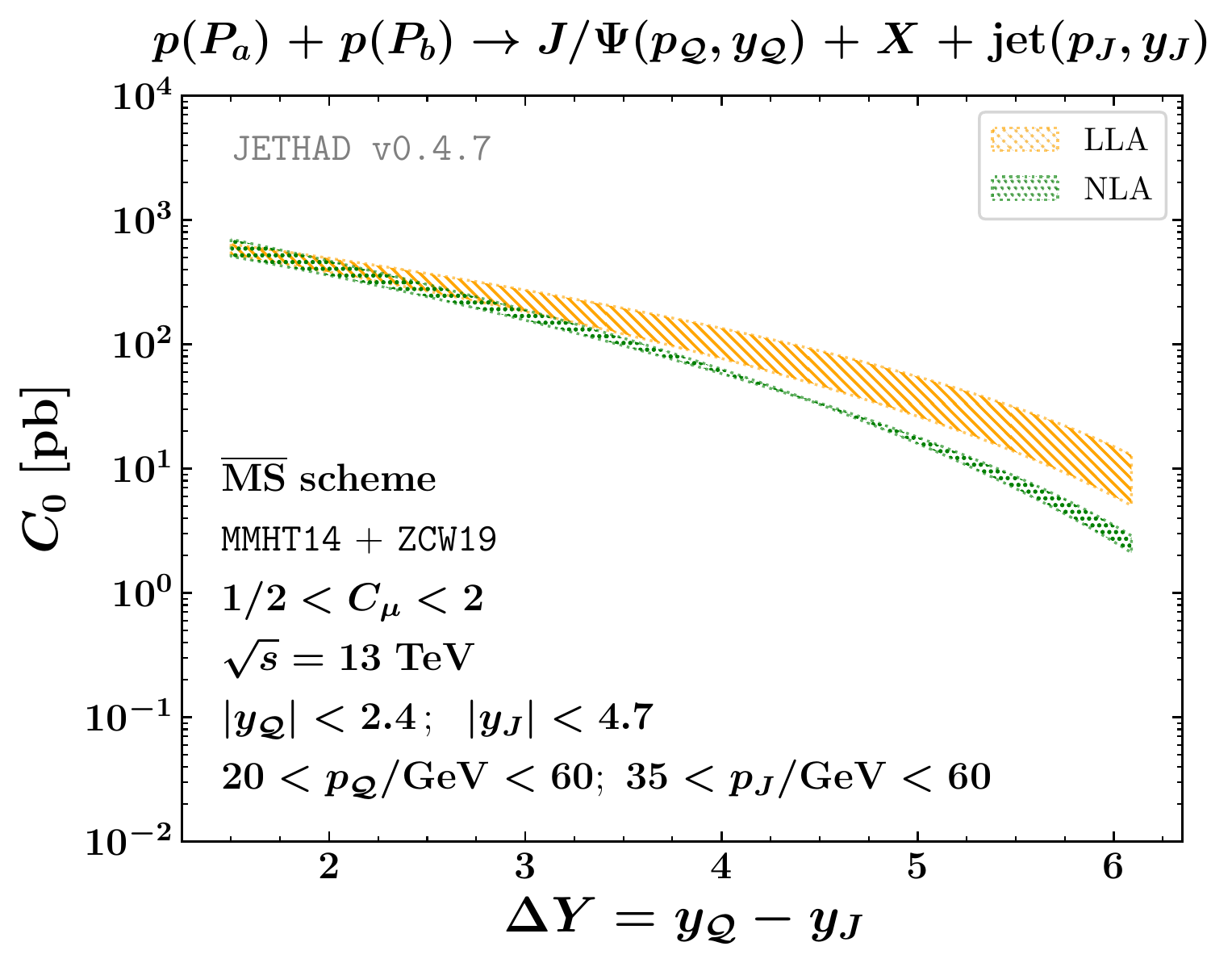}
   \includegraphics[scale=0.50,clip]{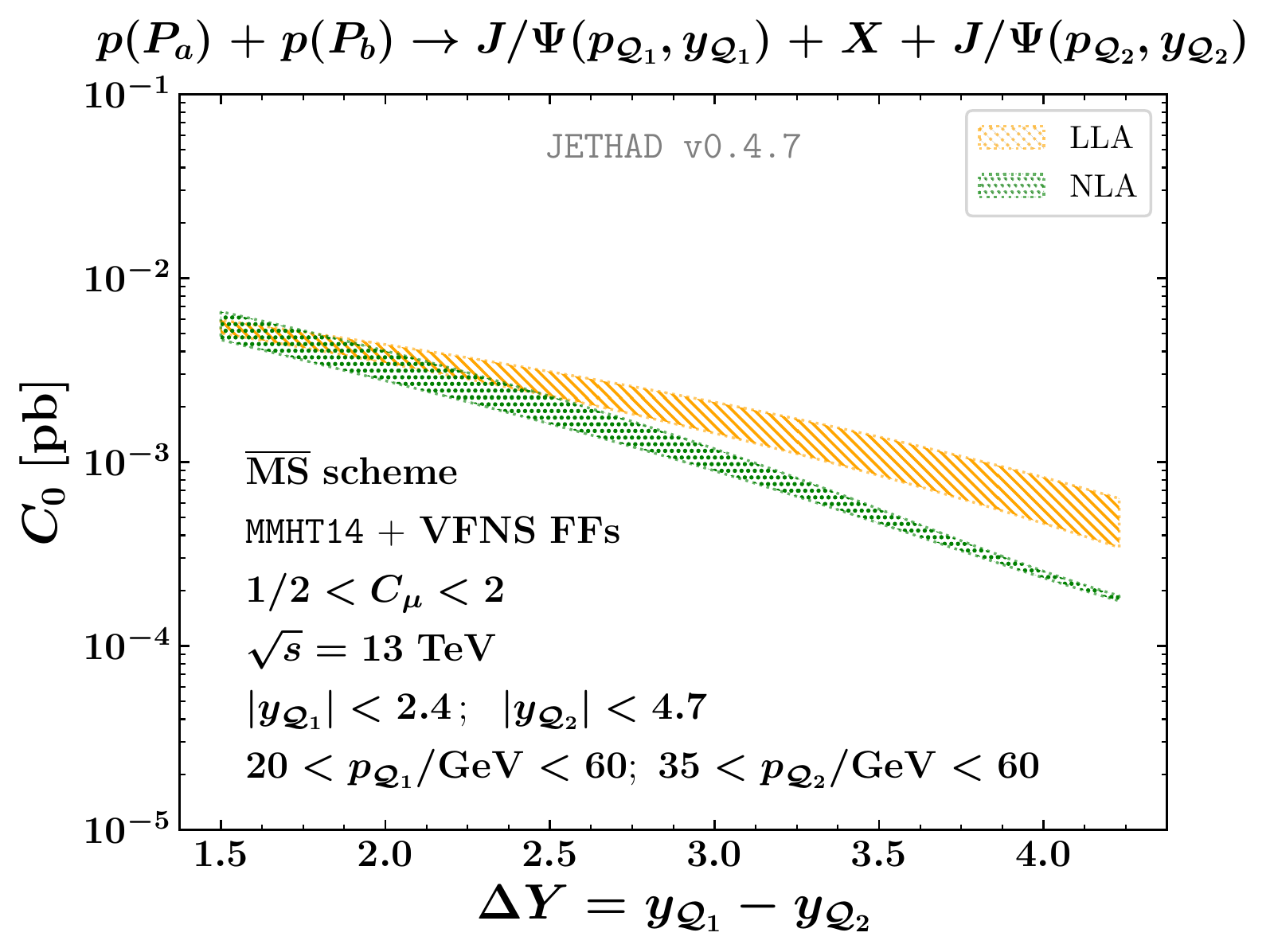}
\caption{$\DY$-distribution in the $\JPsi$~$+$~jet (left) and in the $\JPsi$~$+$~$\JPsi$ (right) channel, at $\sqrt{s} = 13$ TeV. Quarkonium fragmentation is described in terms of {\tt ZCW19} collinear FFs.}
\label{fig:C0_psv}
\end{figure*}

\subsection{$\DY$-distribution}
\label{deltaY}
We start our phe\-no\-me\-no\-lo\-gi\-cal analysis by considering the $\varphi$-summed cross section differential in the $\Delta Y \equiv y_{\cal Q} - y_{J}$ interval or $\DY$-distribution. It is obtained by integrating the ${\cal C}_0$ (see Eq.\eref{Cn_NLA_MSb}) over transverse momenta and rapidities of the two emitted objects, and imposing a fixed value of $\DY$. One has
\begin{eqnarray}
 C_0 &=&
 \int_{y_\Q^{\rm min}}^{y_\Q^{\rm max}} \hspace{-0.15cm} \drv y_\Q
 \int_{y_J^{\rm min}}^{y_J^{\rm max}} \hspace{-0.15cm} \drv y_J
 \int_{p_\Q^{\rm min}}^{p_\Q^{\rm max}} \hspace{-0.15cm} \drv |\vec p_\Q|
 \int_{p_J^{\rm min}}^{p_J^{\rm max}} \hspace{-0.15cm} \drv |\vec p_J|
 \nonumber \\
 &\times&
 \delta (\DY - (y_\Q - y_J))
 \, \,
 {\cal C}_0\left(|\vec p_\Q|, |\vec p_J|, y_\Q, y_J \right)
 \, .
 \label{DY_distribution}
\end{eqnarray}
The light jet is tagged in kinematic configurations typical of current studies at the CMS detector\tcite{Khachatryan:2016udy}, namely~35~GeV~$< p_J <$~60~GeV and $|y_J| < 4.7$. 
The quarkonium transverse momentum is choosen to be in the range 20~GeV~$< |\vec p_{\cal Q}| <$~60~GeV, for consistency with the VFNS treatment. 
Concerning the rapidity range of the quarkonia, we allow the detection to be done only by the CMS barrel detector and not by endcaps, \emph{i.e.}~$|y_{\cal Q}| < 2.4$.

In Fig.\tref{fig:C0_psv} we compare the NLA $\DY$-behavior of $C_0$ with the corresponding prediction at LLA, in the $J/\psi$-plus-jet channel (left) and in the $J/\psi$-plus-$J/\psi$ channel (right). We note that values of $C_0$ are everywhere larger than 0.5 pb in the $J/\psi$-plus-jet channel (left). Although being substantially lower than the one for heavy-baryon and heavy-light meson emissions\tcite{Celiberto:2021dzy,Celiberto:2021fdp}, the statistics is promising.
The downtrend with $\DY$ of our distributions both at LLA and NLA is a common feature of all the hadronic semi-hard reactions investigated so far. Although the high-energy resummation predicts a growth with energy of the partonic-subprocess cross section, its convolution with parent gluon PDFs leads, as a net effect, to a falloff with $\Delta Y$ of both LLA and NLA predictions.
We observe a partial stabilization of the high-energy series, with NLA bands almost overlapped to LLA ones at lower values of $\DY$, and their mutual distance becoming wider in the large-$\DY$ range.
In the left panel of Fig. \ref{fig:Azm} we corroborate our stability claims by carrying out a systematic study under variation of scale. We allow the ratio $C_{\mu}$ to vary between $1$ and $30$, noting that there are no very significant variations, especially at high $\Delta Y$.

\subsection{$\varphi$-distribution}
The second observable that we consider in our phenomenological analysis is the $\varphi$-distribution, or simply azimuthal distribution, defined as
\begin{equation}
\begin{split}
 \frac{1}{\sigma}
 \frac{\drv \sigma}{\drv \varphi} & = \frac{1}{2 \pi} \left\{ 1 + 2 \sum_{n = 1}^{\infty} \cos(n \varphi) \langle \cos(n \varphi) \rangle \right\} \\ &
 = \frac{1}{2 \pi} \left\{ 1 + 2 \sum_{n =1 }^{\infty} \cos(n \varphi) R_{n0} \right\} \; \; ,
\end{split}
\label{dsigma_dphi}
\end{equation}
where $R_{n,0} = C_n / C_0$ (with $C_n$ being the azimuthal coefficient  conformal spin $n$, integrated over the final-state phase space in the same way as in Eq.\eref{DY_distribution}). We checked the numerical stability of our calculation by progressively raising the effective upper limit of the $n$-sum in Eq.\eref{dsigma_dphi}.  An excellent numerical convergence was found at $n_{\rm max} = 20$. We adopt the
same final-state kinematic cuts introduced in subsection \ref{deltaY}.

\begin{figure*}[!t]
\centering
   \hspace{-0.40cm}
   \includegraphics[scale=0.50,clip]{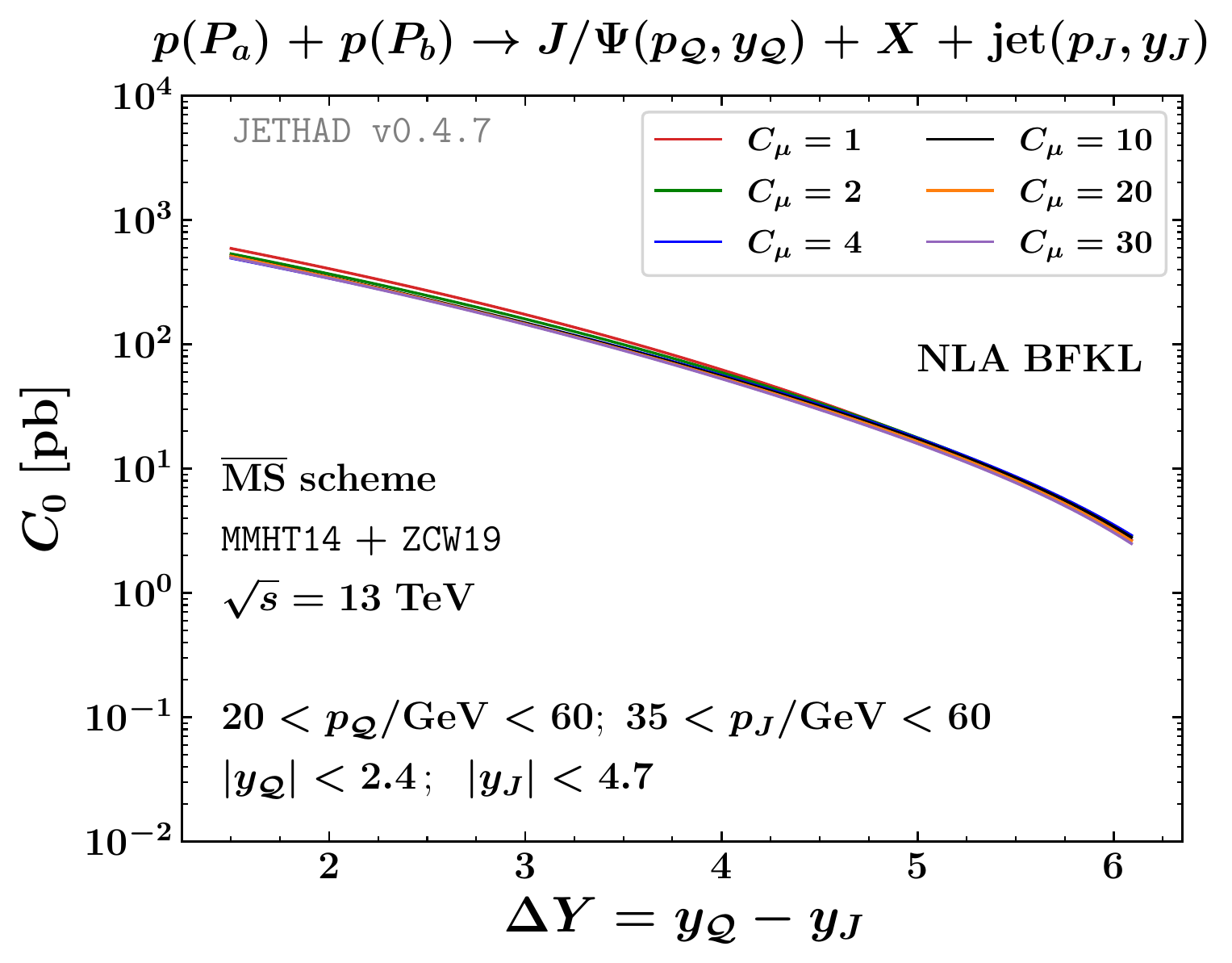}
   \includegraphics[scale=0.50,clip]{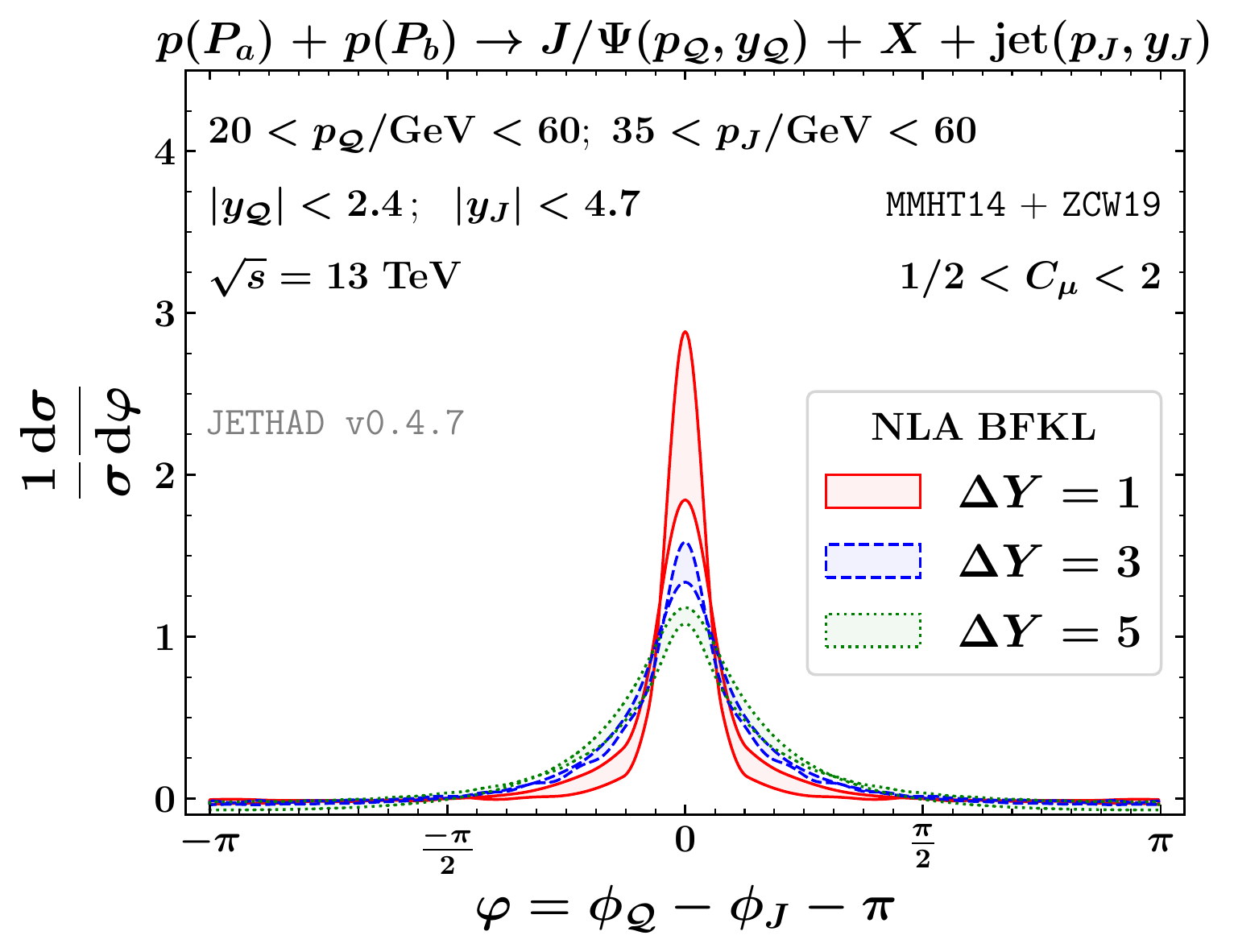}
\caption{In the left panel, $\DY$-distribution in the $\JPsi$~$+$~jet for $C_{\mu}=1,2,4,10,20,30$ at $13$ TeV. In the right panel, NLA predictions for the $\varphi$-distribution in the $\JPsi$~$+$~jet channel, at $\sqrt{s} = 13$ TeV, and for three distinct
values of $\Delta Y$. Quarkonium fragmentation is described in terms of {\tt ZCW19} collinear FFs.}
\label{fig:Azm}
\end{figure*}

In the right panel of Fig.\tref{fig:Azm} we present predictions for the azimuthal distribution as a function of $\varphi$ and for three distinct values of the rapidity interval, $\DY = 1, 3, 5$, in the $\JPsi$~$+$~jet channel. Results were obtained by making use of the {\tt ZCW19} set.

The peculiar behavior of these observables corroborates the assumption that we are probing a regime
where the BFKL treatment is valid. All distributions present a distinct peak at $\varphi = 0$, namely
when the quarkonium and the jet are emitted in back-to-back configurations. When $\Delta Y$ increases, the peak height decreases, while the distribution width broadens. This reflects the fact that larger
rapidity intervals bring to a more significant decorrelation of the quarkonium-jet system, so that the
number of back-to-back events diminish.

\section{Conclusion and outlook}

The general outcomes of results presented in this work are:
\begin{itemize}
    \item[\textbullet] Inclusive forward $J/\psi$ emissions, accompanied by a light-jet emissions lead to a favorable statistics for the $\DY$-differential cross section. Effects of stabilization of the high-energy resummation under higher-order corrections and scale variation are considered. \\
    
    \item[\textbullet] Studying the azimuthal distribution of quarkonium-plus-jet processes around natural values of $\mu_R$ and $\mu_F$ scales is feasible. This observable can be easily measured at the LHC, thus offering us the possibility of doing stringent tests of the high-energy resummation.
\end{itemize}

As anticipated, a more complete analysis including the production channel initiated by gluon (through the {\tt ZCW19$^+$} FF set built in terms of both the constituent heavy-quark FF\tcite{Zheng:2019dfk} and the gluon one\tcite{Braaten:1993rw}) and the study of emissions of the $\Upsilon$, can be found in \cite{Celiberto:2022dyf}. This work represents a further step in our ongoing program on heavy-flavored emissions and the natural continuation is the possible matching between our VFNS analysis and the approach developed in Ref.\tcite{Boussarie:2017oae}.

Remarkably, single-forward vector-quarkonium e\-mis\-sions offer us a peerless opportunity to access the proton structure at low $x$.
In particular, they represents golden channels to study the \emph{unintegrated gluon density} (UGD), whose low-$x$ evolution is controlled by the BFKL equation.
The information on the UGD gathered by quarkonium studies\tcite{Bautista:2016xnp,Garcia:2019tne,Hentschinski:2020yfm} will complement the one already known from deep-inelastic-scattering structure functions\tcite{Hentschinski:2012kr}, light vector-meson helicity amplitudes and cross sections\tcite{Anikin:2009bf,Anikin:2011sa,Besse:2013muy,Bolognino:2018rhb,Bolognino:2018mlw,Bolognino:2019bko,Bolognino:2019pba,Celiberto:2019slj,Bolognino:2021niq,Bolognino:2021gjm,Bolognino:2022uty,Celiberto:2022fam,Bolognino:2022ndh}, and forward Drell--Yan di-lepton distributions\tcite{Motyka:2014lya,Brzeminski:2016lwh,Motyka:2016lta,Celiberto:2018muu}.
Moreover, observables sensitive to quarkonium production are relevant to shed light on the intersection regime between the BFKL UGD and (un)polarized transverse-momentum-dependent gluon densities\tcite{Bacchetta:2020vty,Celiberto:2021zww,Bacchetta:2021oht,Bacchetta:2021lvw,Bacchetta:2021twk,Bacchetta:2022esb,Bacchetta:2022crh,Bacchetta:2022nyv}.

\section*{Acknowledgments}
\label{sec:acknowledgments}

F.G.C. acknowledges support from the INFN/NIN\-PHA project and thanks the Universit\`a degli Studi di Pavia for the warm hospitality.
M.F. acknowledges support from the INFN/QFT@COL\-LI\-DERS project.

\bibliographystyle{apsrev}
\bibliography{references}

\begin{thebibliography}{149}
\expandafter\ifx\csname natexlab\endcsname\relax\def\natexlab#1{#1}\fi
\expandafter\ifx\csname bibnamefont\endcsname\relax
  \def\bibnamefont#1{#1}\fi
\expandafter\ifx\csname bibfnamefont\endcsname\relax
  \def\bibfnamefont#1{#1}\fi
\expandafter\ifx\csname citenamefont\endcsname\relax
  \def\citenamefont#1{#1}\fi
\expandafter\ifx\csname url\endcsname\relax
  \def\url#1{\texttt{#1}}\fi
\expandafter\ifx\csname urlprefix\endcsname\relax\def\urlprefix{URL }\fi
\providecommand{\bibinfo}[2]{#2}
\providecommand{\eprint}[2][]{\url{#2}}

\bibitem[{\citenamefont{Chapon et~al.}(2022)}]{Chapon:2020heu}
\bibinfo{author}{\bibfnamefont{E.}~\bibnamefont{Chapon}} \bibnamefont{et~al.},
  \bibinfo{journal}{Prog. Part. Nucl. Phys.} \textbf{\bibinfo{volume}{122}},
  \bibinfo{pages}{103906} (\bibinfo{year}{2022}), \eprint{2012.14161}.

\bibitem[{\citenamefont{Anchordoqui et~al.}(2022)}]{Anchordoqui:2021ghd}
\bibinfo{author}{\bibfnamefont{L.~A.} \bibnamefont{Anchordoqui}}
  \bibnamefont{et~al.}, \bibinfo{journal}{Phys. Rept.}
  \textbf{\bibinfo{volume}{968}}, \bibinfo{pages}{1} (\bibinfo{year}{2022}),
  \eprint{2109.10905}.

\bibitem[{\citenamefont{Feng et~al.}(2022)}]{Feng:2022inv}
\bibinfo{author}{\bibfnamefont{J.~L.} \bibnamefont{Feng}} \bibnamefont{et~al.}
  (\bibinfo{year}{2022}), \eprint{2203.05090}.

\bibitem[{\citenamefont{Celiberto}(2022{\natexlab{a}})}]{Celiberto:2022rfj}
\bibinfo{author}{\bibfnamefont{F.~G.} \bibnamefont{Celiberto}},
  \bibinfo{journal}{Phys. Rev. D} \textbf{\bibinfo{volume}{105}},
  \bibinfo{pages}{114008} (\bibinfo{year}{2022}{\natexlab{a}}),
  \eprint{2204.06497}.

\bibitem[{\citenamefont{Hentschinski et~al.}(2022)}]{Hentschinski:2022xnd}
\bibinfo{author}{\bibfnamefont{M.}~\bibnamefont{Hentschinski}}
  \bibnamefont{et~al.}, in \emph{\bibinfo{booktitle}{{2022 Snowmass Summer
  Study}}} (\bibinfo{year}{2022}), \eprint{2203.08129}.

\bibitem[{\citenamefont{Accardi et~al.}(2016)}]{Accardi:2012qut}
\bibinfo{author}{\bibfnamefont{A.}~\bibnamefont{Accardi}} \bibnamefont{et~al.},
  \bibinfo{journal}{Eur. Phys. J. A} \textbf{\bibinfo{volume}{52}},
  \bibinfo{pages}{268} (\bibinfo{year}{2016}), \eprint{1212.1701}.

\bibitem[{\citenamefont{Abdul~Khalek et~al.}(2021)}]{AbdulKhalek:2021gbh}
\bibinfo{author}{\bibfnamefont{R.}~\bibnamefont{Abdul~Khalek}}
  \bibnamefont{et~al.} (\bibinfo{year}{2021}), \eprint{2103.05419}.

\bibitem[{\citenamefont{Abdul~Khalek et~al.}(2022)}]{Khalek:2022bzd}
\bibinfo{author}{\bibfnamefont{R.}~\bibnamefont{Abdul~Khalek}}
  \bibnamefont{et~al.}, in \emph{\bibinfo{booktitle}{{2022 Snowmass Summer
  Study}}} (\bibinfo{year}{2022}), \eprint{2203.13199}.

\bibitem[{\citenamefont{Acosta et~al.}(2022)\citenamefont{Acosta, Barberis,
  Hurley, Li, Colin, Wood, and Zuo}}]{Acosta:2022ejc}
\bibinfo{author}{\bibfnamefont{D.}~\bibnamefont{Acosta}},
  \bibinfo{author}{\bibfnamefont{E.}~\bibnamefont{Barberis}},
  \bibinfo{author}{\bibfnamefont{N.}~\bibnamefont{Hurley}},
  \bibinfo{author}{\bibfnamefont{W.}~\bibnamefont{Li}},
  \bibinfo{author}{\bibfnamefont{O.~M.} \bibnamefont{Colin}},
  \bibinfo{author}{\bibfnamefont{D.}~\bibnamefont{Wood}}, \bibnamefont{and}
  \bibinfo{author}{\bibfnamefont{X.}~\bibnamefont{Zuo}}, in
  \emph{\bibinfo{booktitle}{{2022 Snowmass Summer Study}}}
  (\bibinfo{year}{2022}), \eprint{2203.06258}.

\bibitem[{\citenamefont{Adachi et~al.}(2022)}]{AlexanderAryshev:2022pkx}
\bibinfo{author}{\bibfnamefont{I.}~\bibnamefont{Adachi}} \bibnamefont{et~al.}
  (\bibinfo{collaboration}{ILC International Development Team and ILC
  Community}) (\bibinfo{year}{2022}), \eprint{2203.07622}.

\bibitem[{\citenamefont{Celiberto
  et~al.}(2016{\natexlab{a}})\citenamefont{Celiberto, Jenkovszky, and
  Myronenko}}]{Celiberto:2016yek}
\bibinfo{author}{\bibfnamefont{F.~G.} \bibnamefont{Celiberto}},
  \bibinfo{author}{\bibfnamefont{L.}~\bibnamefont{Jenkovszky}},
  \bibnamefont{and}
  \bibinfo{author}{\bibfnamefont{V.}~\bibnamefont{Myronenko}},
  \bibinfo{journal}{EPJ Web Conf.} \textbf{\bibinfo{volume}{125}},
  \bibinfo{pages}{04012} (\bibinfo{year}{2016}{\natexlab{a}}),
  \eprint{1608.04646}.

\bibitem[{\citenamefont{Celiberto
  et~al.}(2017{\natexlab{a}})\citenamefont{Celiberto, Fiore, and
  Jenkovszky}}]{Celiberto:2016jse}
\bibinfo{author}{\bibfnamefont{F.~G.} \bibnamefont{Celiberto}},
  \bibinfo{author}{\bibfnamefont{R.}~\bibnamefont{Fiore}}, \bibnamefont{and}
  \bibinfo{author}{\bibfnamefont{L.}~\bibnamefont{Jenkovszky}},
  \bibinfo{journal}{AIP Conf. Proc.} \textbf{\bibinfo{volume}{1819}},
  \bibinfo{pages}{030005} (\bibinfo{year}{2017}{\natexlab{a}}),
  \eprint{1612.00797}.

\bibitem[{\citenamefont{Brunner et~al.}(2022)}]{Brunner:2022usy}
\bibinfo{author}{\bibfnamefont{O.}~\bibnamefont{Brunner}} \bibnamefont{et~al.}
  (\bibinfo{year}{2022}), \eprint{2203.09186}.

\bibitem[{\citenamefont{Arbuzov et~al.}(2021)}]{Arbuzov:2020cqg}
\bibinfo{author}{\bibfnamefont{A.}~\bibnamefont{Arbuzov}} \bibnamefont{et~al.},
  \bibinfo{journal}{Prog. Part. Nucl. Phys.} \textbf{\bibinfo{volume}{119}},
  \bibinfo{pages}{103858} (\bibinfo{year}{2021}), \eprint{2011.15005}.

\bibitem[{\citenamefont{Abazov et~al.}(2021)}]{Abazov:2021hku}
\bibinfo{author}{\bibfnamefont{V.~M.} \bibnamefont{Abazov}}
  \bibnamefont{et~al.} (\bibinfo{collaboration}{SPD proto})
  (\bibinfo{year}{2021}), \eprint{2102.00442}.

\bibitem[{\citenamefont{Bernardi et~al.}(2022)}]{Bernardi:2022hny}
\bibinfo{author}{\bibfnamefont{G.}~\bibnamefont{Bernardi}} \bibnamefont{et~al.}
  (\bibinfo{year}{2022}), \eprint{2203.06520}.

\bibitem[{\citenamefont{Amoroso et~al.}(2022)}]{Amoroso:2022eow}
\bibinfo{author}{\bibfnamefont{S.}~\bibnamefont{Amoroso}} \bibnamefont{et~al.},
  in \emph{\bibinfo{booktitle}{{2022 Snowmass Summer Study}}}
  (\bibinfo{year}{2022}), \eprint{2203.13923}.

\bibitem[{\citenamefont{Celiberto
  et~al.}(2021{\natexlab{a}})\citenamefont{Celiberto, Fucilla, Ivanov,
  Mohammed, and Papa}}]{Celiberto:2018hdy}
\bibinfo{author}{\bibfnamefont{F.~G.} \bibnamefont{Celiberto}},
  \bibinfo{author}{\bibfnamefont{M.}~\bibnamefont{Fucilla}},
  \bibinfo{author}{\bibfnamefont{D.~{\relax Yu}.} \bibnamefont{Ivanov}},
  \bibinfo{author}{\bibfnamefont{M.~M.~A.} \bibnamefont{Mohammed}},
  \bibnamefont{and} \bibinfo{author}{\bibfnamefont{A.}~\bibnamefont{Papa}}, in
  \emph{\bibinfo{booktitle}{{2022 Snowmass Summer Study}}}
  (\bibinfo{year}{2021}{\natexlab{a}}).

\bibitem[{\citenamefont{Klein et~al.}(2020)}]{Klein:2020nvu}
\bibinfo{author}{\bibfnamefont{S.}~\bibnamefont{Klein}} \bibnamefont{et~al.}
  (\bibinfo{year}{2020}), \eprint{2009.03838}.

\bibitem[{\citenamefont{Canepa and D'Onofrio}(2022)}]{2064676}
\bibinfo{author}{\bibfnamefont{A.}~\bibnamefont{Canepa}} \bibnamefont{and}
  \bibinfo{author}{\bibfnamefont{M.}~\bibnamefont{D'Onofrio}}
  (\bibinfo{year}{2022}).

\bibitem[{\citenamefont{de~Blas et~al.}(2022)}]{MuonCollider:2022xlm}
\bibinfo{author}{\bibfnamefont{J.}~\bibnamefont{de~Blas}} \bibnamefont{et~al.}
  (\bibinfo{collaboration}{Muon Collider}) (\bibinfo{year}{2022}),
  \eprint{2203.07261}.

\bibitem[{\citenamefont{Aim\`e et~al.}(2022)}]{Aime:2022flm}
\bibinfo{author}{\bibfnamefont{C.}~\bibnamefont{Aim\`e}} \bibnamefont{et~al.}
  (\bibinfo{year}{2022}), \eprint{2203.07256}.

\bibitem[{\citenamefont{Gribov et~al.}(1983)\citenamefont{Gribov, Levin, and
  Ryskin}}]{Gribov:1983ivg}
\bibinfo{author}{\bibfnamefont{L.~V.} \bibnamefont{Gribov}},
  \bibinfo{author}{\bibfnamefont{E.~M.} \bibnamefont{Levin}}, \bibnamefont{and}
  \bibinfo{author}{\bibfnamefont{M.~G.} \bibnamefont{Ryskin}},
  \bibinfo{journal}{Phys. Rept.} \textbf{\bibinfo{volume}{100}},
  \bibinfo{pages}{1} (\bibinfo{year}{1983}).

\bibitem[{\citenamefont{Celiberto}(2017)}]{Celiberto:2017ius}
\bibinfo{author}{\bibfnamefont{F.~G.} \bibnamefont{Celiberto}}, Ph.D. thesis,
  \bibinfo{school}{Universit\`a della Calabria and INFN-Cosenza}
  (\bibinfo{year}{2017}), \eprint{1707.04315}.

\bibitem[{\citenamefont{Fadin et~al.}(1975)\citenamefont{Fadin, Kuraev, and
  Lipatov}}]{Fadin:1975cb}
\bibinfo{author}{\bibfnamefont{V.~S.} \bibnamefont{Fadin}},
  \bibinfo{author}{\bibfnamefont{E.}~\bibnamefont{Kuraev}}, \bibnamefont{and}
  \bibinfo{author}{\bibfnamefont{L.}~\bibnamefont{Lipatov}},
  \bibinfo{journal}{Phys. Lett. B} \textbf{\bibinfo{volume}{60}},
  \bibinfo{pages}{50} (\bibinfo{year}{1975}).

\bibitem[{\citenamefont{Kuraev et~al.}(1976)\citenamefont{Kuraev, Lipatov, and
  Fadin}}]{Kuraev:1976ge}
\bibinfo{author}{\bibfnamefont{E.~A.} \bibnamefont{Kuraev}},
  \bibinfo{author}{\bibfnamefont{L.~N.} \bibnamefont{Lipatov}},
  \bibnamefont{and} \bibinfo{author}{\bibfnamefont{V.~S.} \bibnamefont{Fadin}},
  \bibinfo{journal}{Sov. Phys. JETP} \textbf{\bibinfo{volume}{44}},
  \bibinfo{pages}{443} (\bibinfo{year}{1976}).

\bibitem[{\citenamefont{Kuraev et~al.}(1977)\citenamefont{Kuraev, Lipatov, and
  Fadin}}]{Kuraev:1977fs}
\bibinfo{author}{\bibfnamefont{E.}~\bibnamefont{Kuraev}},
  \bibinfo{author}{\bibfnamefont{L.}~\bibnamefont{Lipatov}}, \bibnamefont{and}
  \bibinfo{author}{\bibfnamefont{V.~S.} \bibnamefont{Fadin}},
  \bibinfo{journal}{Sov.\ Phys.\ JETP} \textbf{\bibinfo{volume}{45}},
  \bibinfo{pages}{199} (\bibinfo{year}{1977}).

\bibitem[{\citenamefont{Balitsky and Lipatov}(1978)}]{Balitsky:1978ic}
\bibinfo{author}{\bibfnamefont{I.}~\bibnamefont{Balitsky}} \bibnamefont{and}
  \bibinfo{author}{\bibfnamefont{L.}~\bibnamefont{Lipatov}},
  \bibinfo{journal}{Sov.\ J.\ Nucl.\ Phys.} \textbf{\bibinfo{volume}{28}},
  \bibinfo{pages}{822} (\bibinfo{year}{1978}).

\bibitem[{\citenamefont{Fadin and Lipatov}(1998)}]{Fadin:1998py}
\bibinfo{author}{\bibfnamefont{V.~S.} \bibnamefont{Fadin}} \bibnamefont{and}
  \bibinfo{author}{\bibfnamefont{L.~N.} \bibnamefont{Lipatov}},
  \bibinfo{journal}{Phys. Lett. B} \textbf{\bibinfo{volume}{429}},
  \bibinfo{pages}{127} (\bibinfo{year}{1998}), \eprint{hep-ph/9802290}.

\bibitem[{\citenamefont{Ciafaloni and Camici}(1998)}]{Ciafaloni:1998gs}
\bibinfo{author}{\bibfnamefont{M.}~\bibnamefont{Ciafaloni}} \bibnamefont{and}
  \bibinfo{author}{\bibfnamefont{G.}~\bibnamefont{Camici}},
  \bibinfo{journal}{Phys. Lett. B} \textbf{\bibinfo{volume}{430}},
  \bibinfo{pages}{349} (\bibinfo{year}{1998}), \eprint{hep-ph/9803389}.

\bibitem[{\citenamefont{Fadin et~al.}(1999)\citenamefont{Fadin, Fiore, and
  Papa}}]{Fadin:1998jv}
\bibinfo{author}{\bibfnamefont{V.~S.} \bibnamefont{Fadin}},
  \bibinfo{author}{\bibfnamefont{R.}~\bibnamefont{Fiore}}, \bibnamefont{and}
  \bibinfo{author}{\bibfnamefont{A.}~\bibnamefont{Papa}},
  \bibinfo{journal}{Phys. Rev. D} \textbf{\bibinfo{volume}{60}},
  \bibinfo{pages}{074025} (\bibinfo{year}{1999}), \eprint{hep-ph/9812456}.

\bibitem[{\citenamefont{Fadin and
  Gorbachev}(2000{\natexlab{a}})}]{Fadin:2000kx}
\bibinfo{author}{\bibfnamefont{V.~S.} \bibnamefont{Fadin}} \bibnamefont{and}
  \bibinfo{author}{\bibfnamefont{D.~A.} \bibnamefont{Gorbachev}},
  \bibinfo{journal}{JETP Lett.} \textbf{\bibinfo{volume}{71}},
  \bibinfo{pages}{222} (\bibinfo{year}{2000}{\natexlab{a}}).

\bibitem[{\citenamefont{Fadin and
  Gorbachev}(2000{\natexlab{b}})}]{Fadin:2000hu}
\bibinfo{author}{\bibfnamefont{V.~S.} \bibnamefont{Fadin}} \bibnamefont{and}
  \bibinfo{author}{\bibfnamefont{D.~A.} \bibnamefont{Gorbachev}},
  \bibinfo{journal}{Phys. Atom. Nucl.} \textbf{\bibinfo{volume}{63}},
  \bibinfo{pages}{2157} (\bibinfo{year}{2000}{\natexlab{b}}).

\bibitem[{\citenamefont{Fadin and Fiore}(2005{\natexlab{a}})}]{Fadin:2004zq}
\bibinfo{author}{\bibfnamefont{V.~S.} \bibnamefont{Fadin}} \bibnamefont{and}
  \bibinfo{author}{\bibfnamefont{R.}~\bibnamefont{Fiore}},
  \bibinfo{journal}{Phys. Lett. B} \textbf{\bibinfo{volume}{610}},
  \bibinfo{pages}{61} (\bibinfo{year}{2005}{\natexlab{a}}),
  \bibinfo{note}{[Erratum: Phys.Lett.B 621, 320 (2005)]},
  \eprint{hep-ph/0412386}.

\bibitem[{\citenamefont{Fadin and Fiore}(2005{\natexlab{b}})}]{Fadin:2005zj}
\bibinfo{author}{\bibfnamefont{V.~S.} \bibnamefont{Fadin}} \bibnamefont{and}
  \bibinfo{author}{\bibfnamefont{R.}~\bibnamefont{Fiore}},
  \bibinfo{journal}{Phys. Rev. D} \textbf{\bibinfo{volume}{72}},
  \bibinfo{pages}{014018} (\bibinfo{year}{2005}{\natexlab{b}}),
  \eprint{hep-ph/0502045}.

\bibitem[{\citenamefont{Mueller and Navelet}(1987)}]{Mueller:1986ey}
\bibinfo{author}{\bibfnamefont{A.~H.} \bibnamefont{Mueller}} \bibnamefont{and}
  \bibinfo{author}{\bibfnamefont{H.}~\bibnamefont{Navelet}},
  \bibinfo{journal}{Nucl. Phys. B} \textbf{\bibinfo{volume}{282}},
  \bibinfo{pages}{727} (\bibinfo{year}{1987}).

\bibitem[{\citenamefont{Colferai et~al.}(2010)\citenamefont{Colferai,
  Schwennsen, Szymanowski, and Wallon}}]{Colferai:2010wu}
\bibinfo{author}{\bibfnamefont{D.}~\bibnamefont{Colferai}},
  \bibinfo{author}{\bibfnamefont{F.}~\bibnamefont{Schwennsen}},
  \bibinfo{author}{\bibfnamefont{L.}~\bibnamefont{Szymanowski}},
  \bibnamefont{and} \bibinfo{author}{\bibfnamefont{S.}~\bibnamefont{Wallon}},
  \bibinfo{journal}{JHEP} \textbf{\bibinfo{volume}{12}}, \bibinfo{pages}{026}
  (\bibinfo{year}{2010}), \eprint{1002.1365}.

\bibitem[{\citenamefont{Caporale
  et~al.}(2013{\natexlab{a}})\citenamefont{Caporale, Ivanov, Murdaca, and
  Papa}}]{Caporale:2012ih}
\bibinfo{author}{\bibfnamefont{F.}~\bibnamefont{Caporale}},
  \bibinfo{author}{\bibfnamefont{D.}~\bibnamefont{Ivanov}},
  \bibinfo{author}{\bibfnamefont{B.}~\bibnamefont{Murdaca}}, \bibnamefont{and}
  \bibinfo{author}{\bibfnamefont{A.}~\bibnamefont{Papa}},
  \bibinfo{journal}{Nucl. Phys. B} \textbf{\bibinfo{volume}{877}},
  \bibinfo{pages}{73} (\bibinfo{year}{2013}{\natexlab{a}}), \eprint{1211.7225}.

\bibitem[{\citenamefont{Duclou\'e et~al.}(2013)\citenamefont{Duclou\'e,
  Szymanowski, and Wallon}}]{Ducloue:2013hia}
\bibinfo{author}{\bibfnamefont{B.}~\bibnamefont{Duclou\'e}},
  \bibinfo{author}{\bibfnamefont{L.}~\bibnamefont{Szymanowski}},
  \bibnamefont{and} \bibinfo{author}{\bibfnamefont{S.}~\bibnamefont{Wallon}},
  \bibinfo{journal}{JHEP} \textbf{\bibinfo{volume}{05}}, \bibinfo{pages}{096}
  (\bibinfo{year}{2013}), \eprint{1302.7012}.

\bibitem[{\citenamefont{Duclou\'e et~al.}(2014)\citenamefont{Duclou\'e,
  Szymanowski, and Wallon}}]{Ducloue:2013bva}
\bibinfo{author}{\bibfnamefont{B.}~\bibnamefont{Duclou\'e}},
  \bibinfo{author}{\bibfnamefont{L.}~\bibnamefont{Szymanowski}},
  \bibnamefont{and} \bibinfo{author}{\bibfnamefont{S.}~\bibnamefont{Wallon}},
  \bibinfo{journal}{Phys. Rev. Lett.} \textbf{\bibinfo{volume}{112}},
  \bibinfo{pages}{082003} (\bibinfo{year}{2014}), \eprint{1309.3229}.

\bibitem[{\citenamefont{Caporale
  et~al.}(2013{\natexlab{b}})\citenamefont{Caporale, Murdaca, Sabio~Vera, and
  Salas}}]{Caporale:2013uva}
\bibinfo{author}{\bibfnamefont{F.}~\bibnamefont{Caporale}},
  \bibinfo{author}{\bibfnamefont{B.}~\bibnamefont{Murdaca}},
  \bibinfo{author}{\bibfnamefont{A.}~\bibnamefont{Sabio~Vera}},
  \bibnamefont{and} \bibinfo{author}{\bibfnamefont{C.}~\bibnamefont{Salas}},
  \bibinfo{journal}{Nucl. Phys. B} \textbf{\bibinfo{volume}{875}},
  \bibinfo{pages}{134} (\bibinfo{year}{2013}{\natexlab{b}}),
  \eprint{1305.4620}.

\bibitem[{\citenamefont{Caporale et~al.}(2014)\citenamefont{Caporale, Ivanov,
  Murdaca, and Papa}}]{Caporale:2014gpa}
\bibinfo{author}{\bibfnamefont{F.}~\bibnamefont{Caporale}},
  \bibinfo{author}{\bibfnamefont{D.~{\relax Yu}.} \bibnamefont{Ivanov}},
  \bibinfo{author}{\bibfnamefont{B.}~\bibnamefont{Murdaca}}, \bibnamefont{and}
  \bibinfo{author}{\bibfnamefont{A.}~\bibnamefont{Papa}},
  \bibinfo{journal}{Eur. Phys. J. C} \textbf{\bibinfo{volume}{74}},
  \bibinfo{pages}{3084} (\bibinfo{year}{2014}), \bibinfo{note}{[Erratum:
  Eur.Phys.J.C 75, 535 (2015)]}, \eprint{1407.8431}.

\bibitem[{\citenamefont{Colferai and Niccoli}(2015)}]{Colferai:2015zfa}
\bibinfo{author}{\bibfnamefont{D.}~\bibnamefont{Colferai}} \bibnamefont{and}
  \bibinfo{author}{\bibfnamefont{A.}~\bibnamefont{Niccoli}},
  \bibinfo{journal}{JHEP} \textbf{\bibinfo{volume}{04}}, \bibinfo{pages}{071}
  (\bibinfo{year}{2015}), \eprint{1501.07442}.

\bibitem[{\citenamefont{Caporale et~al.}(2015)\citenamefont{Caporale, Ivanov,
  Murdaca, and Papa}}]{Caporale:2015uva}
\bibinfo{author}{\bibfnamefont{F.}~\bibnamefont{Caporale}},
  \bibinfo{author}{\bibfnamefont{D.~{\relax Yu}.} \bibnamefont{Ivanov}},
  \bibinfo{author}{\bibfnamefont{B.}~\bibnamefont{Murdaca}}, \bibnamefont{and}
  \bibinfo{author}{\bibfnamefont{A.}~\bibnamefont{Papa}},
  \bibinfo{journal}{Phys. Rev. D} \textbf{\bibinfo{volume}{91}},
  \bibinfo{pages}{114009} (\bibinfo{year}{2015}), \eprint{1504.06471}.

\bibitem[{\citenamefont{Duclou\'e et~al.}(2015)\citenamefont{Duclou\'e,
  Szymanowski, and Wallon}}]{Ducloue:2015jba}
\bibinfo{author}{\bibfnamefont{B.}~\bibnamefont{Duclou\'e}},
  \bibinfo{author}{\bibfnamefont{L.}~\bibnamefont{Szymanowski}},
  \bibnamefont{and} \bibinfo{author}{\bibfnamefont{S.}~\bibnamefont{Wallon}},
  \bibinfo{journal}{Phys. Rev. D} \textbf{\bibinfo{volume}{92}},
  \bibinfo{pages}{076002} (\bibinfo{year}{2015}), \eprint{1507.04735}.

\bibitem[{\citenamefont{Celiberto
  et~al.}(2015{\natexlab{a}})\citenamefont{Celiberto, Ivanov, Murdaca, and
  Papa}}]{Celiberto:2015yba}
\bibinfo{author}{\bibfnamefont{F.~G.} \bibnamefont{Celiberto}},
  \bibinfo{author}{\bibfnamefont{D.~{\relax Yu}.} \bibnamefont{Ivanov}},
  \bibinfo{author}{\bibfnamefont{B.}~\bibnamefont{Murdaca}}, \bibnamefont{and}
  \bibinfo{author}{\bibfnamefont{A.}~\bibnamefont{Papa}},
  \bibinfo{journal}{Eur. Phys. J. C} \textbf{\bibinfo{volume}{75}},
  \bibinfo{pages}{292} (\bibinfo{year}{2015}{\natexlab{a}}),
  \eprint{1504.08233}.

\bibitem[{\citenamefont{Celiberto
  et~al.}(2015{\natexlab{b}})\citenamefont{Celiberto, Ivanov, Murdaca, and
  Papa}}]{Celiberto:2015mpa}
\bibinfo{author}{\bibfnamefont{F.~G.} \bibnamefont{Celiberto}},
  \bibinfo{author}{\bibfnamefont{D.~{\relax Yu}.} \bibnamefont{Ivanov}},
  \bibinfo{author}{\bibfnamefont{B.}~\bibnamefont{Murdaca}}, \bibnamefont{and}
  \bibinfo{author}{\bibfnamefont{A.}~\bibnamefont{Papa}},
  \bibinfo{journal}{Acta Phys. Polon. Supp.} \textbf{\bibinfo{volume}{8}},
  \bibinfo{pages}{935} (\bibinfo{year}{2015}{\natexlab{b}}),
  \eprint{1510.01626}.

\bibitem[{\citenamefont{Celiberto
  et~al.}(2016{\natexlab{b}})\citenamefont{Celiberto, Ivanov, Murdaca, and
  Papa}}]{Celiberto:2016ygs}
\bibinfo{author}{\bibfnamefont{F.~G.} \bibnamefont{Celiberto}},
  \bibinfo{author}{\bibfnamefont{D.~{\relax Yu}.} \bibnamefont{Ivanov}},
  \bibinfo{author}{\bibfnamefont{B.}~\bibnamefont{Murdaca}}, \bibnamefont{and}
  \bibinfo{author}{\bibfnamefont{A.}~\bibnamefont{Papa}},
  \bibinfo{journal}{Eur. Phys. J. C} \textbf{\bibinfo{volume}{76}},
  \bibinfo{pages}{224} (\bibinfo{year}{2016}{\natexlab{b}}),
  \eprint{1601.07847}.

\bibitem[{\citenamefont{Celiberto
  et~al.}(2016{\natexlab{c}})\citenamefont{Celiberto, Ivanov, Murdaca, and
  Papa}}]{Celiberto:2016vva}
\bibinfo{author}{\bibfnamefont{F.~G.} \bibnamefont{Celiberto}},
  \bibinfo{author}{\bibfnamefont{D.~{\relax Yu}.} \bibnamefont{Ivanov}},
  \bibinfo{author}{\bibfnamefont{B.}~\bibnamefont{Murdaca}}, \bibnamefont{and}
  \bibinfo{author}{\bibfnamefont{A.}~\bibnamefont{Papa}},
  \bibinfo{journal}{PoS} \textbf{\bibinfo{volume}{DIS2016}},
  \bibinfo{pages}{176} (\bibinfo{year}{2016}{\natexlab{c}}),
  \eprint{1606.08892}.

\bibitem[{\citenamefont{Caporale et~al.}(2018)\citenamefont{Caporale,
  Celiberto, Chachamis, Gordo~G\'omez, and Sabio~Vera}}]{Caporale:2018qnm}
\bibinfo{author}{\bibfnamefont{F.}~\bibnamefont{Caporale}},
  \bibinfo{author}{\bibfnamefont{F.~G.} \bibnamefont{Celiberto}},
  \bibinfo{author}{\bibfnamefont{G.}~\bibnamefont{Chachamis}},
  \bibinfo{author}{\bibfnamefont{D.}~\bibnamefont{Gordo~G\'omez}},
  \bibnamefont{and}
  \bibinfo{author}{\bibfnamefont{A.}~\bibnamefont{Sabio~Vera}},
  \bibinfo{journal}{Nucl. Phys. B} \textbf{\bibinfo{volume}{935}},
  \bibinfo{pages}{412} (\bibinfo{year}{2018}), \eprint{1806.06309}.

\bibitem[{\citenamefont{de~Le\'on et~al.}(2021)\citenamefont{de~Le\'on,
  Chachamis, and Sabio~Vera}}]{deLeon:2021ecb}
\bibinfo{author}{\bibfnamefont{N.~B.} \bibnamefont{de~Le\'on}},
  \bibinfo{author}{\bibfnamefont{G.}~\bibnamefont{Chachamis}},
  \bibnamefont{and}
  \bibinfo{author}{\bibfnamefont{A.}~\bibnamefont{Sabio~Vera}},
  \bibinfo{journal}{Eur. Phys. J. C} \textbf{\bibinfo{volume}{81}},
  \bibinfo{pages}{1019} (\bibinfo{year}{2021}), \eprint{2106.11255}.

\bibitem[{\citenamefont{Celiberto and Papa}(2022)}]{Celiberto:2022gji}
\bibinfo{author}{\bibfnamefont{F.~G.} \bibnamefont{Celiberto}}
  \bibnamefont{and} \bibinfo{author}{\bibfnamefont{A.}~\bibnamefont{Papa}}
  (\bibinfo{year}{2022}), \eprint{2207.05015}.

\bibitem[{\citenamefont{Celiberto
  et~al.}(2016{\natexlab{d}})\citenamefont{Celiberto, Ivanov, Murdaca, and
  Papa}}]{Celiberto:2016hae}
\bibinfo{author}{\bibfnamefont{F.~G.} \bibnamefont{Celiberto}},
  \bibinfo{author}{\bibfnamefont{D.~{\relax Yu}.} \bibnamefont{Ivanov}},
  \bibinfo{author}{\bibfnamefont{B.}~\bibnamefont{Murdaca}}, \bibnamefont{and}
  \bibinfo{author}{\bibfnamefont{A.}~\bibnamefont{Papa}},
  \bibinfo{journal}{Phys. Rev. D} \textbf{\bibinfo{volume}{94}},
  \bibinfo{pages}{034013} (\bibinfo{year}{2016}{\natexlab{d}}),
  \eprint{1604.08013}.

\bibitem[{\citenamefont{Celiberto
  et~al.}(2017{\natexlab{b}})\citenamefont{Celiberto, Ivanov, Murdaca, and
  Papa}}]{Celiberto:2016zgb}
\bibinfo{author}{\bibfnamefont{F.~G.} \bibnamefont{Celiberto}},
  \bibinfo{author}{\bibfnamefont{D.~{\relax Yu}.} \bibnamefont{Ivanov}},
  \bibinfo{author}{\bibfnamefont{B.}~\bibnamefont{Murdaca}}, \bibnamefont{and}
  \bibinfo{author}{\bibfnamefont{A.}~\bibnamefont{Papa}}, \bibinfo{journal}{AIP
  Conf. Proc.} \textbf{\bibinfo{volume}{1819}}, \bibinfo{pages}{060005}
  (\bibinfo{year}{2017}{\natexlab{b}}), \eprint{1611.04811}.

\bibitem[{\citenamefont{Celiberto
  et~al.}(2017{\natexlab{c}})\citenamefont{Celiberto, Ivanov, Murdaca, and
  Papa}}]{Celiberto:2017ptm}
\bibinfo{author}{\bibfnamefont{F.~G.} \bibnamefont{Celiberto}},
  \bibinfo{author}{\bibfnamefont{D.~{\relax Yu}.} \bibnamefont{Ivanov}},
  \bibinfo{author}{\bibfnamefont{B.}~\bibnamefont{Murdaca}}, \bibnamefont{and}
  \bibinfo{author}{\bibfnamefont{A.}~\bibnamefont{Papa}},
  \bibinfo{journal}{Eur. Phys. J. C} \textbf{\bibinfo{volume}{77}},
  \bibinfo{pages}{382} (\bibinfo{year}{2017}{\natexlab{c}}),
  \eprint{1701.05077}.

\bibitem[{\citenamefont{Celiberto
  et~al.}(2017{\natexlab{d}})\citenamefont{Celiberto, Ivanov, Murdaca, and
  Papa}}]{Celiberto:2017uae}
\bibinfo{author}{\bibfnamefont{F.~G.} \bibnamefont{Celiberto}},
  \bibinfo{author}{\bibfnamefont{D.~{\relax Yu}.} \bibnamefont{Ivanov}},
  \bibinfo{author}{\bibfnamefont{B.}~\bibnamefont{Murdaca}}, \bibnamefont{and}
  \bibinfo{author}{\bibfnamefont{A.}~\bibnamefont{Papa}}, in
  \emph{\bibinfo{booktitle}{{25th Low-x Meeting}}}
  (\bibinfo{year}{2017}{\natexlab{d}}), \eprint{1709.01128}.

\bibitem[{\citenamefont{Celiberto
  et~al.}(2017{\natexlab{e}})\citenamefont{Celiberto, Ivanov, Murdaca, and
  Papa}}]{Celiberto:2017ydk}
\bibinfo{author}{\bibfnamefont{F.~G.} \bibnamefont{Celiberto}},
  \bibinfo{author}{\bibfnamefont{D.~{\relax Yu}.} \bibnamefont{Ivanov}},
  \bibinfo{author}{\bibfnamefont{B.}~\bibnamefont{Murdaca}}, \bibnamefont{and}
  \bibinfo{author}{\bibfnamefont{A.}~\bibnamefont{Papa}}, in
  \emph{\bibinfo{booktitle}{{17th conference on Elastic and Diffractive
  Scattering}}} (\bibinfo{year}{2017}{\natexlab{e}}), \eprint{1709.04758}.

\bibitem[{\citenamefont{Caporale
  et~al.}(2016{\natexlab{a}})\citenamefont{Caporale, Chachamis, Murdaca, and
  Sabio~Vera}}]{Caporale:2015vya}
\bibinfo{author}{\bibfnamefont{F.}~\bibnamefont{Caporale}},
  \bibinfo{author}{\bibfnamefont{G.}~\bibnamefont{Chachamis}},
  \bibinfo{author}{\bibfnamefont{B.}~\bibnamefont{Murdaca}}, \bibnamefont{and}
  \bibinfo{author}{\bibfnamefont{A.}~\bibnamefont{Sabio~Vera}},
  \bibinfo{journal}{Phys. Rev. Lett.} \textbf{\bibinfo{volume}{116}},
  \bibinfo{pages}{012001} (\bibinfo{year}{2016}{\natexlab{a}}),
  \eprint{1508.07711}.

\bibitem[{\citenamefont{Caporale
  et~al.}(2016{\natexlab{b}})\citenamefont{Caporale, Celiberto, Chachamis, and
  Sabio~Vera}}]{Caporale:2015int}
\bibinfo{author}{\bibfnamefont{F.}~\bibnamefont{Caporale}},
  \bibinfo{author}{\bibfnamefont{F.~G.} \bibnamefont{Celiberto}},
  \bibinfo{author}{\bibfnamefont{G.}~\bibnamefont{Chachamis}},
  \bibnamefont{and}
  \bibinfo{author}{\bibfnamefont{A.}~\bibnamefont{Sabio~Vera}},
  \bibinfo{journal}{Eur. Phys. J. C} \textbf{\bibinfo{volume}{76}},
  \bibinfo{pages}{165} (\bibinfo{year}{2016}{\natexlab{b}}),
  \eprint{1512.03364}.

\bibitem[{\citenamefont{Caporale
  et~al.}(2016{\natexlab{c}})\citenamefont{Caporale, Celiberto, Chachamis,
  Gordo~G\'omez, and Sabio~Vera}}]{Caporale:2016soq}
\bibinfo{author}{\bibfnamefont{F.}~\bibnamefont{Caporale}},
  \bibinfo{author}{\bibfnamefont{F.~G.} \bibnamefont{Celiberto}},
  \bibinfo{author}{\bibfnamefont{G.}~\bibnamefont{Chachamis}},
  \bibinfo{author}{\bibfnamefont{D.}~\bibnamefont{Gordo~G\'omez}},
  \bibnamefont{and}
  \bibinfo{author}{\bibfnamefont{A.}~\bibnamefont{Sabio~Vera}},
  \bibinfo{journal}{Nucl. Phys. B} \textbf{\bibinfo{volume}{910}},
  \bibinfo{pages}{374} (\bibinfo{year}{2016}{\natexlab{c}}),
  \eprint{1603.07785}.

\bibitem[{\citenamefont{Caporale
  et~al.}(2016{\natexlab{d}})\citenamefont{Caporale, Celiberto, Chachamis, and
  Sabio~Vera}}]{Caporale:2016vxt}
\bibinfo{author}{\bibfnamefont{F.}~\bibnamefont{Caporale}},
  \bibinfo{author}{\bibfnamefont{F.~G.} \bibnamefont{Celiberto}},
  \bibinfo{author}{\bibfnamefont{G.}~\bibnamefont{Chachamis}},
  \bibnamefont{and}
  \bibinfo{author}{\bibfnamefont{A.}~\bibnamefont{Sabio~Vera}},
  \bibinfo{journal}{PoS} \textbf{\bibinfo{volume}{DIS2016}},
  \bibinfo{pages}{177} (\bibinfo{year}{2016}{\natexlab{d}}),
  \eprint{1610.01880}.

\bibitem[{\citenamefont{Caporale
  et~al.}(2017{\natexlab{a}})\citenamefont{Caporale, Celiberto, Chachamis,
  Gordo~G\'omez, and Sabio~Vera}}]{Caporale:2016xku}
\bibinfo{author}{\bibfnamefont{F.}~\bibnamefont{Caporale}},
  \bibinfo{author}{\bibfnamefont{F.~G.} \bibnamefont{Celiberto}},
  \bibinfo{author}{\bibfnamefont{G.}~\bibnamefont{Chachamis}},
  \bibinfo{author}{\bibfnamefont{D.}~\bibnamefont{Gordo~G\'omez}},
  \bibnamefont{and}
  \bibinfo{author}{\bibfnamefont{A.}~\bibnamefont{Sabio~Vera}},
  \bibinfo{journal}{Eur. Phys. J. C} \textbf{\bibinfo{volume}{77}},
  \bibinfo{pages}{5} (\bibinfo{year}{2017}{\natexlab{a}}), \eprint{1606.00574}.

\bibitem[{\citenamefont{Celiberto}(2016)}]{Celiberto:2016vhn}
\bibinfo{author}{\bibfnamefont{F.~G.} \bibnamefont{Celiberto}},
  \bibinfo{journal}{Frascati Phys. Ser.} \textbf{\bibinfo{volume}{63}},
  \bibinfo{pages}{43} (\bibinfo{year}{2016}), \eprint{1606.07327}.

\bibitem[{\citenamefont{Caporale
  et~al.}(2017{\natexlab{b}})\citenamefont{Caporale, Celiberto, Chachamis,
  Gordo~G\'omez, and Sabio~Vera}}]{Caporale:2016djm}
\bibinfo{author}{\bibfnamefont{F.}~\bibnamefont{Caporale}},
  \bibinfo{author}{\bibfnamefont{F.~G.} \bibnamefont{Celiberto}},
  \bibinfo{author}{\bibfnamefont{G.}~\bibnamefont{Chachamis}},
  \bibinfo{author}{\bibfnamefont{D.}~\bibnamefont{Gordo~G\'omez}},
  \bibnamefont{and}
  \bibinfo{author}{\bibfnamefont{A.}~\bibnamefont{Sabio~Vera}},
  \bibinfo{journal}{AIP Conf. Proc.} \textbf{\bibinfo{volume}{1819}},
  \bibinfo{pages}{060009} (\bibinfo{year}{2017}{\natexlab{b}}),
  \eprint{1611.04813}.

\bibitem[{\citenamefont{Caporale
  et~al.}(2017{\natexlab{c}})\citenamefont{Caporale, Celiberto, Chachamis,
  Gordo~Gomez, Murdaca, and Sabio~Vera}}]{Caporale:2016pqe}
\bibinfo{author}{\bibfnamefont{F.}~\bibnamefont{Caporale}},
  \bibinfo{author}{\bibfnamefont{F.~G.} \bibnamefont{Celiberto}},
  \bibinfo{author}{\bibfnamefont{G.}~\bibnamefont{Chachamis}},
  \bibinfo{author}{\bibfnamefont{D.}~\bibnamefont{Gordo~Gomez}},
  \bibinfo{author}{\bibfnamefont{B.}~\bibnamefont{Murdaca}}, \bibnamefont{and}
  \bibinfo{author}{\bibfnamefont{A.}~\bibnamefont{Sabio~Vera}}, in
  \emph{\bibinfo{booktitle}{{24th Low-x Meeting}}}
  (\bibinfo{year}{2017}{\natexlab{c}}), vol.~\bibinfo{volume}{5},
  p.~\bibinfo{pages}{47}, \eprint{1610.04765}.

\bibitem[{\citenamefont{Chachamis
  et~al.}(2016{\natexlab{a}})\citenamefont{Chachamis, Caporale, Celiberto,
  Gomez~Gordo, and Sabio~Vera}}]{Chachamis:2016qct}
\bibinfo{author}{\bibfnamefont{G.}~\bibnamefont{Chachamis}},
  \bibinfo{author}{\bibfnamefont{F.}~\bibnamefont{Caporale}},
  \bibinfo{author}{\bibfnamefont{F.}~\bibnamefont{Celiberto}},
  \bibinfo{author}{\bibfnamefont{D.}~\bibnamefont{Gomez~Gordo}},
  \bibnamefont{and}
  \bibinfo{author}{\bibfnamefont{A.}~\bibnamefont{Sabio~Vera}},
  \bibinfo{journal}{PoS} \textbf{\bibinfo{volume}{DIS2016}},
  \bibinfo{pages}{178} (\bibinfo{year}{2016}{\natexlab{a}}).

\bibitem[{\citenamefont{Chachamis
  et~al.}(2016{\natexlab{b}})\citenamefont{Chachamis, Caporale, Celiberto,
  Gomez, and Sabio~Vera}}]{Chachamis:2016lyi}
\bibinfo{author}{\bibfnamefont{G.}~\bibnamefont{Chachamis}},
  \bibinfo{author}{\bibfnamefont{F.}~\bibnamefont{Caporale}},
  \bibinfo{author}{\bibfnamefont{F.~G.} \bibnamefont{Celiberto}},
  \bibinfo{author}{\bibfnamefont{D.~G.} \bibnamefont{Gomez}}, \bibnamefont{and}
  \bibinfo{author}{\bibfnamefont{A.}~\bibnamefont{Sabio~Vera}}
  (\bibinfo{year}{2016}{\natexlab{b}}), \eprint{1610.01342}.

\bibitem[{\citenamefont{Caporale
  et~al.}(2017{\natexlab{d}})\citenamefont{Caporale, Celiberto, Chachamis,
  Gordo~G\'omez, and Sabio~Vera}}]{Caporale:2016lnh}
\bibinfo{author}{\bibfnamefont{F.}~\bibnamefont{Caporale}},
  \bibinfo{author}{\bibfnamefont{F.~G.} \bibnamefont{Celiberto}},
  \bibinfo{author}{\bibfnamefont{G.}~\bibnamefont{Chachamis}},
  \bibinfo{author}{\bibfnamefont{D.}~\bibnamefont{Gordo~G\'omez}},
  \bibnamefont{and}
  \bibinfo{author}{\bibfnamefont{A.}~\bibnamefont{Sabio~Vera}},
  \bibinfo{journal}{EPJ Web Conf.} \textbf{\bibinfo{volume}{164}},
  \bibinfo{pages}{07027} (\bibinfo{year}{2017}{\natexlab{d}}),
  \eprint{1612.02771}.

\bibitem[{\citenamefont{Caporale
  et~al.}(2017{\natexlab{e}})\citenamefont{Caporale, Celiberto, Chachamis,
  Gordo~G\'omez, and Sabio~Vera}}]{Caporale:2016zkc}
\bibinfo{author}{\bibfnamefont{F.}~\bibnamefont{Caporale}},
  \bibinfo{author}{\bibfnamefont{F.~G.} \bibnamefont{Celiberto}},
  \bibinfo{author}{\bibfnamefont{G.}~\bibnamefont{Chachamis}},
  \bibinfo{author}{\bibfnamefont{D.}~\bibnamefont{Gordo~G\'omez}},
  \bibnamefont{and}
  \bibinfo{author}{\bibfnamefont{A.}~\bibnamefont{Sabio~Vera}},
  \bibinfo{journal}{Phys. Rev. D} \textbf{\bibinfo{volume}{95}},
  \bibinfo{pages}{074007} (\bibinfo{year}{2017}{\natexlab{e}}),
  \eprint{1612.05428}.

\bibitem[{\citenamefont{Chachamis et~al.}(2018)\citenamefont{Chachamis,
  Caporale, Celiberto, Gordo~Gomez, and Sabio~Vera}}]{Chachamis:2017vfa}
\bibinfo{author}{\bibfnamefont{G.}~\bibnamefont{Chachamis}},
  \bibinfo{author}{\bibfnamefont{F.}~\bibnamefont{Caporale}},
  \bibinfo{author}{\bibfnamefont{F.~G.} \bibnamefont{Celiberto}},
  \bibinfo{author}{\bibfnamefont{D.}~\bibnamefont{Gordo~Gomez}},
  \bibnamefont{and}
  \bibinfo{author}{\bibfnamefont{A.}~\bibnamefont{Sabio~Vera}},
  \bibinfo{journal}{PoS} \textbf{\bibinfo{volume}{DIS2017}},
  \bibinfo{pages}{067} (\bibinfo{year}{2018}), \eprint{1709.02649}.

\bibitem[{\citenamefont{Caporale
  et~al.}(2017{\natexlab{f}})\citenamefont{Caporale, Celiberto, Gordo~Gomez,
  Sabio~Vera, and Chachamis}}]{Caporale:2017jqj}
\bibinfo{author}{\bibfnamefont{F.}~\bibnamefont{Caporale}},
  \bibinfo{author}{\bibfnamefont{F.~G.} \bibnamefont{Celiberto}},
  \bibinfo{author}{\bibfnamefont{D.}~\bibnamefont{Gordo~Gomez}},
  \bibinfo{author}{\bibfnamefont{A.}~\bibnamefont{Sabio~Vera}},
  \bibnamefont{and}
  \bibinfo{author}{\bibfnamefont{G.}~\bibnamefont{Chachamis}}, in
  \emph{\bibinfo{booktitle}{{25th Low-x Meeting}}}
  (\bibinfo{year}{2017}{\natexlab{f}}), \eprint{1801.00014}.

\bibitem[{\citenamefont{Bolognino
  et~al.}(2018{\natexlab{a}})\citenamefont{Bolognino, Celiberto, Ivanov,
  Mohammed, and Papa}}]{Bolognino:2018oth}
\bibinfo{author}{\bibfnamefont{A.~D.} \bibnamefont{Bolognino}},
  \bibinfo{author}{\bibfnamefont{F.~G.} \bibnamefont{Celiberto}},
  \bibinfo{author}{\bibfnamefont{D.~{\relax Yu}.} \bibnamefont{Ivanov}},
  \bibinfo{author}{\bibfnamefont{M.~M.} \bibnamefont{Mohammed}},
  \bibnamefont{and} \bibinfo{author}{\bibfnamefont{A.}~\bibnamefont{Papa}},
  \bibinfo{journal}{Eur. Phys. J. C} \textbf{\bibinfo{volume}{78}},
  \bibinfo{pages}{772} (\bibinfo{year}{2018}{\natexlab{a}}),
  \eprint{1808.05483}.

\bibitem[{\citenamefont{Bolognino
  et~al.}(2019{\natexlab{a}})\citenamefont{Bolognino, Celiberto, Ivanov,
  Mohammed, and Papa}}]{Bolognino:2019cac}
\bibinfo{author}{\bibfnamefont{A.~D.} \bibnamefont{Bolognino}},
  \bibinfo{author}{\bibfnamefont{F.~G.} \bibnamefont{Celiberto}},
  \bibinfo{author}{\bibfnamefont{D.~{\relax Yu}.} \bibnamefont{Ivanov}},
  \bibinfo{author}{\bibfnamefont{M.~M.~A.} \bibnamefont{Mohammed}},
  \bibnamefont{and} \bibinfo{author}{\bibfnamefont{A.}~\bibnamefont{Papa}},
  \bibinfo{journal}{PoS} \textbf{\bibinfo{volume}{DIS2019}},
  \bibinfo{pages}{049} (\bibinfo{year}{2019}{\natexlab{a}}),
  \eprint{1906.11800}.

\bibitem[{\citenamefont{Bolognino
  et~al.}(2019{\natexlab{b}})\citenamefont{Bolognino, Celiberto, Ivanov,
  Mohammed, and Papa}}]{Bolognino:2019yqj}
\bibinfo{author}{\bibfnamefont{A.~D.} \bibnamefont{Bolognino}},
  \bibinfo{author}{\bibfnamefont{F.~G.} \bibnamefont{Celiberto}},
  \bibinfo{author}{\bibfnamefont{D.~{\relax Yu}.} \bibnamefont{Ivanov}},
  \bibinfo{author}{\bibfnamefont{M.~M.} \bibnamefont{Mohammed}},
  \bibnamefont{and} \bibinfo{author}{\bibfnamefont{A.}~\bibnamefont{Papa}},
  \bibinfo{journal}{Acta Phys. Polon. Supp.} \textbf{\bibinfo{volume}{12}},
  \bibinfo{pages}{773} (\bibinfo{year}{2019}{\natexlab{b}}),
  \eprint{1902.04511}.

\bibitem[{\citenamefont{Celiberto et~al.}(2020)\citenamefont{Celiberto, Ivanov,
  and Papa}}]{Celiberto:2020rxb}
\bibinfo{author}{\bibfnamefont{F.~G.} \bibnamefont{Celiberto}},
  \bibinfo{author}{\bibfnamefont{D.~{\relax Yu}.} \bibnamefont{Ivanov}},
  \bibnamefont{and} \bibinfo{author}{\bibfnamefont{A.}~\bibnamefont{Papa}},
  \bibinfo{journal}{Phys. Rev. D} \textbf{\bibinfo{volume}{102}},
  \bibinfo{pages}{094019} (\bibinfo{year}{2020}), \eprint{2008.10513}.

\bibitem[{\citenamefont{Celiberto
  et~al.}(2021{\natexlab{b}})\citenamefont{Celiberto, Ivanov, Mohammed, and
  Papa}}]{Celiberto:2020tmb}
\bibinfo{author}{\bibfnamefont{F.~G.} \bibnamefont{Celiberto}},
  \bibinfo{author}{\bibfnamefont{D.~{\relax Yu}.} \bibnamefont{Ivanov}},
  \bibinfo{author}{\bibfnamefont{M.~M.~A.} \bibnamefont{Mohammed}},
  \bibnamefont{and} \bibinfo{author}{\bibfnamefont{A.}~\bibnamefont{Papa}},
  \bibinfo{journal}{Eur. Phys. J. C} \textbf{\bibinfo{volume}{81}},
  \bibinfo{pages}{293} (\bibinfo{year}{2021}{\natexlab{b}}),
  \eprint{2008.00501}.

\bibitem[{\citenamefont{Celiberto
  et~al.}(2022{\natexlab{a}})\citenamefont{Celiberto, Ivanov, Mohammed, and
  Papa}}]{Celiberto:2021fjf}
\bibinfo{author}{\bibfnamefont{F.~G.} \bibnamefont{Celiberto}},
  \bibinfo{author}{\bibfnamefont{D.~Y.} \bibnamefont{Ivanov}},
  \bibinfo{author}{\bibfnamefont{M.~M.~A.} \bibnamefont{Mohammed}},
  \bibnamefont{and} \bibinfo{author}{\bibfnamefont{A.}~\bibnamefont{Papa}},
  \bibinfo{journal}{SciPost Phys. Proc.} \textbf{\bibinfo{volume}{8}},
  \bibinfo{pages}{039} (\bibinfo{year}{2022}{\natexlab{a}}),
  \eprint{2107.13037}.

\bibitem[{\citenamefont{Celiberto
  et~al.}(2022{\natexlab{b}})\citenamefont{Celiberto, Fucilla, Papa, Ivanov,
  and Mohammed}}]{Celiberto:2021tky}
\bibinfo{author}{\bibfnamefont{F.~G.} \bibnamefont{Celiberto}},
  \bibinfo{author}{\bibfnamefont{M.}~\bibnamefont{Fucilla}},
  \bibinfo{author}{\bibfnamefont{A.}~\bibnamefont{Papa}},
  \bibinfo{author}{\bibfnamefont{D.~{\relax Yu}.} \bibnamefont{Ivanov}},
  \bibnamefont{and} \bibinfo{author}{\bibfnamefont{M.~M.~A.}
  \bibnamefont{Mohammed}}, \bibinfo{journal}{PoS}
  \textbf{\bibinfo{volume}{EPS-HEP2021}}, \bibinfo{pages}{589}
  (\bibinfo{year}{2022}{\natexlab{b}}), \eprint{2110.09358}.

\bibitem[{\citenamefont{Celiberto
  et~al.}(2022{\natexlab{c}})\citenamefont{Celiberto, Fucilla, Ivanov,
  Mohammed, and Papa}}]{Celiberto:2021txb}
\bibinfo{author}{\bibfnamefont{F.~G.} \bibnamefont{Celiberto}},
  \bibinfo{author}{\bibfnamefont{M.}~\bibnamefont{Fucilla}},
  \bibinfo{author}{\bibfnamefont{D.~{\relax Yu}.} \bibnamefont{Ivanov}},
  \bibinfo{author}{\bibfnamefont{M.~M.~A.} \bibnamefont{Mohammed}},
  \bibnamefont{and} \bibinfo{author}{\bibfnamefont{A.}~\bibnamefont{Papa}},
  \bibinfo{journal}{PoS} \textbf{\bibinfo{volume}{PANIC2021}},
  \bibinfo{pages}{352} (\bibinfo{year}{2022}{\natexlab{c}}),
  \eprint{2111.13090}.

\bibitem[{\citenamefont{Celiberto
  et~al.}(2021{\natexlab{c}})\citenamefont{Celiberto, Fucilla, Ivanov,
  Mohammed, and Papa}}]{Celiberto:2021xpm}
\bibinfo{author}{\bibfnamefont{F.~G.} \bibnamefont{Celiberto}},
  \bibinfo{author}{\bibfnamefont{M.}~\bibnamefont{Fucilla}},
  \bibinfo{author}{\bibfnamefont{D.~{\relax Yu}.} \bibnamefont{Ivanov}},
  \bibinfo{author}{\bibfnamefont{M.~M.~A.} \bibnamefont{Mohammed}},
  \bibnamefont{and} \bibinfo{author}{\bibfnamefont{A.}~\bibnamefont{Papa}}
  (\bibinfo{year}{2021}{\natexlab{c}}), \eprint{2110.12649}.

\bibitem[{\citenamefont{Celiberto
  et~al.}(2022{\natexlab{d}})\citenamefont{Celiberto, Fucilla, Ivanov,
  Mohammed, and Papa}}]{Celiberto:2022fgx}
\bibinfo{author}{\bibfnamefont{F.~G.} \bibnamefont{Celiberto}},
  \bibinfo{author}{\bibfnamefont{M.}~\bibnamefont{Fucilla}},
  \bibinfo{author}{\bibfnamefont{D.~Y.} \bibnamefont{Ivanov}},
  \bibinfo{author}{\bibfnamefont{M.~M.~A.} \bibnamefont{Mohammed}},
  \bibnamefont{and} \bibinfo{author}{\bibfnamefont{A.}~\bibnamefont{Papa}},
  \bibinfo{journal}{JHEP} \textbf{\bibinfo{volume}{08}}, \bibinfo{pages}{092}
  (\bibinfo{year}{2022}{\natexlab{d}}), \eprint{2205.02681}.

\bibitem[{\citenamefont{Golec-Biernat et~al.}(2018)\citenamefont{Golec-Biernat,
  Motyka, and Stebel}}]{Golec-Biernat:2018kem}
\bibinfo{author}{\bibfnamefont{K.}~\bibnamefont{Golec-Biernat}},
  \bibinfo{author}{\bibfnamefont{L.}~\bibnamefont{Motyka}}, \bibnamefont{and}
  \bibinfo{author}{\bibfnamefont{T.}~\bibnamefont{Stebel}},
  \bibinfo{journal}{JHEP} \textbf{\bibinfo{volume}{12}}, \bibinfo{pages}{091}
  (\bibinfo{year}{2018}), \eprint{1811.04361}.

\bibitem[{\citenamefont{Boussarie et~al.}(2018)\citenamefont{Boussarie,
  Duclou\'e, Szymanowski, and Wallon}}]{Boussarie:2017oae}
\bibinfo{author}{\bibfnamefont{R.}~\bibnamefont{Boussarie}},
  \bibinfo{author}{\bibfnamefont{B.}~\bibnamefont{Duclou\'e}},
  \bibinfo{author}{\bibfnamefont{L.}~\bibnamefont{Szymanowski}},
  \bibnamefont{and} \bibinfo{author}{\bibfnamefont{S.}~\bibnamefont{Wallon}},
  \bibinfo{journal}{Phys. Rev. D} \textbf{\bibinfo{volume}{97}},
  \bibinfo{pages}{014008} (\bibinfo{year}{2018}), \eprint{1709.01380}.

\bibitem[{\citenamefont{Celiberto
  et~al.}(2018{\natexlab{a}})\citenamefont{Celiberto, Ivanov, Murdaca, and
  Papa}}]{Celiberto:2017nyx}
\bibinfo{author}{\bibfnamefont{F.~G.} \bibnamefont{Celiberto}},
  \bibinfo{author}{\bibfnamefont{D.~{\relax Yu}.} \bibnamefont{Ivanov}},
  \bibinfo{author}{\bibfnamefont{B.}~\bibnamefont{Murdaca}}, \bibnamefont{and}
  \bibinfo{author}{\bibfnamefont{A.}~\bibnamefont{Papa}},
  \bibinfo{journal}{Phys. Lett. B} \textbf{\bibinfo{volume}{777}},
  \bibinfo{pages}{141} (\bibinfo{year}{2018}{\natexlab{a}}),
  \eprint{1709.10032}.

\bibitem[{\citenamefont{Bolognino
  et~al.}(2019{\natexlab{c}})\citenamefont{Bolognino, Celiberto, Fucilla,
  Ivanov, Murdaca, and Papa}}]{Bolognino:2019ouc}
\bibinfo{author}{\bibfnamefont{A.~D.} \bibnamefont{Bolognino}},
  \bibinfo{author}{\bibfnamefont{F.~G.} \bibnamefont{Celiberto}},
  \bibinfo{author}{\bibfnamefont{M.}~\bibnamefont{Fucilla}},
  \bibinfo{author}{\bibfnamefont{D.~{\relax Yu}.} \bibnamefont{Ivanov}},
  \bibinfo{author}{\bibfnamefont{B.}~\bibnamefont{Murdaca}}, \bibnamefont{and}
  \bibinfo{author}{\bibfnamefont{A.}~\bibnamefont{Papa}},
  \bibinfo{journal}{PoS} \textbf{\bibinfo{volume}{DIS2019}},
  \bibinfo{pages}{067} (\bibinfo{year}{2019}{\natexlab{c}}),
  \eprint{1906.05940}.

\bibitem[{\citenamefont{Bolognino
  et~al.}(2019{\natexlab{d}})\citenamefont{Bolognino, Celiberto, Fucilla,
  Ivanov, and Papa}}]{Bolognino:2019yls}
\bibinfo{author}{\bibfnamefont{A.~D.} \bibnamefont{Bolognino}},
  \bibinfo{author}{\bibfnamefont{F.~G.} \bibnamefont{Celiberto}},
  \bibinfo{author}{\bibfnamefont{M.}~\bibnamefont{Fucilla}},
  \bibinfo{author}{\bibfnamefont{D.~{\relax Yu}.} \bibnamefont{Ivanov}},
  \bibnamefont{and} \bibinfo{author}{\bibfnamefont{A.}~\bibnamefont{Papa}},
  \bibinfo{journal}{Eur. Phys. J. C} \textbf{\bibinfo{volume}{79}},
  \bibinfo{pages}{939} (\bibinfo{year}{2019}{\natexlab{d}}),
  \eprint{1909.03068}.

\bibitem[{\citenamefont{Bolognino
  et~al.}(2019{\natexlab{e}})\citenamefont{Bolognino, Celiberto, Fucilla,
  Ivanov, Murdaca, and Papa}}]{Bolognino:2019ccd}
\bibinfo{author}{\bibfnamefont{A.~D.} \bibnamefont{Bolognino}},
  \bibinfo{author}{\bibfnamefont{F.~G.} \bibnamefont{Celiberto}},
  \bibinfo{author}{\bibfnamefont{M.}~\bibnamefont{Fucilla}},
  \bibinfo{author}{\bibfnamefont{D.~{\relax Yu}.} \bibnamefont{Ivanov}},
  \bibinfo{author}{\bibfnamefont{B.}~\bibnamefont{Murdaca}}, \bibnamefont{and}
  \bibinfo{author}{\bibfnamefont{A.}~\bibnamefont{Papa}},
  \bibinfo{journal}{PoS} \textbf{\bibinfo{volume}{DIS2019}},
  \bibinfo{pages}{067} (\bibinfo{year}{2019}{\natexlab{e}}),
  \eprint{1906.05940}.

\bibitem[{\citenamefont{Celiberto
  et~al.}(2021{\natexlab{d}})\citenamefont{Celiberto, Fucilla, Ivanov, and
  Papa}}]{Celiberto:2021dzy}
\bibinfo{author}{\bibfnamefont{F.~G.} \bibnamefont{Celiberto}},
  \bibinfo{author}{\bibfnamefont{M.}~\bibnamefont{Fucilla}},
  \bibinfo{author}{\bibfnamefont{D.~{\relax Yu}.} \bibnamefont{Ivanov}},
  \bibnamefont{and} \bibinfo{author}{\bibfnamefont{A.}~\bibnamefont{Papa}},
  \bibinfo{journal}{Eur. Phys. J. C} \textbf{\bibinfo{volume}{81}},
  \bibinfo{pages}{780} (\bibinfo{year}{2021}{\natexlab{d}}),
  \eprint{2105.06432}.

\bibitem[{\citenamefont{Celiberto
  et~al.}(2021{\natexlab{e}})\citenamefont{Celiberto, Fucilla, Ivanov,
  Mohammed, and Papa}}]{Celiberto:2021fdp}
\bibinfo{author}{\bibfnamefont{F.~G.} \bibnamefont{Celiberto}},
  \bibinfo{author}{\bibfnamefont{M.}~\bibnamefont{Fucilla}},
  \bibinfo{author}{\bibfnamefont{D.~{\relax Yu}.} \bibnamefont{Ivanov}},
  \bibinfo{author}{\bibfnamefont{M.~M.~A.} \bibnamefont{Mohammed}},
  \bibnamefont{and} \bibinfo{author}{\bibfnamefont{A.}~\bibnamefont{Papa}},
  \bibinfo{journal}{Phys. Rev. D} \textbf{\bibinfo{volume}{104}},
  \bibinfo{pages}{114007} (\bibinfo{year}{2021}{\natexlab{e}}),
  \eprint{2109.11875}.

\bibitem[{\citenamefont{Bolognino
  et~al.}(2022{\natexlab{a}})\citenamefont{Bolognino, Celiberto, Fucilla,
  Ivanov, and Papa}}]{Bolognino:2021zco}
\bibinfo{author}{\bibfnamefont{A.~D.} \bibnamefont{Bolognino}},
  \bibinfo{author}{\bibfnamefont{F.~G.} \bibnamefont{Celiberto}},
  \bibinfo{author}{\bibfnamefont{M.}~\bibnamefont{Fucilla}},
  \bibinfo{author}{\bibfnamefont{D.~{\relax Yu}.} \bibnamefont{Ivanov}},
  \bibnamefont{and} \bibinfo{author}{\bibfnamefont{A.}~\bibnamefont{Papa}},
  \bibinfo{journal}{PoS} \textbf{\bibinfo{volume}{EPS-HEP2021}},
  \bibinfo{pages}{389} (\bibinfo{year}{2022}{\natexlab{a}}).

\bibitem[{\citenamefont{Bolognino
  et~al.}(2022{\natexlab{b}})\citenamefont{Bolognino, Celiberto, Fucilla,
  Ivanov, and Papa}}]{Bolognino:2022wgl}
\bibinfo{author}{\bibfnamefont{A.~D.} \bibnamefont{Bolognino}},
  \bibinfo{author}{\bibfnamefont{F.~G.} \bibnamefont{Celiberto}},
  \bibinfo{author}{\bibfnamefont{M.}~\bibnamefont{Fucilla}},
  \bibinfo{author}{\bibfnamefont{D.~{\relax Yu}.} \bibnamefont{Ivanov}},
  \bibnamefont{and} \bibinfo{author}{\bibfnamefont{A.}~\bibnamefont{Papa}},
  \bibinfo{journal}{PoS} \textbf{\bibinfo{volume}{EPS-HEP2021}},
  \bibinfo{pages}{389} (\bibinfo{year}{2022}{\natexlab{b}}),
  \eprint{2110.12772}.

\bibitem[{\citenamefont{Celiberto and Fucilla}(2022)}]{Celiberto:2022dyf}
\bibinfo{author}{\bibfnamefont{F.~G.} \bibnamefont{Celiberto}}
  \bibnamefont{and} \bibinfo{author}{\bibfnamefont{M.}~\bibnamefont{Fucilla}},
  \bibinfo{journal}{under revision in Eur. Phys. J. C}  (\bibinfo{year}{2022}),
  \eprint{2202.12227}.

\bibitem[{\citenamefont{Celiberto}(2022{\natexlab{b}})}]{Celiberto:2022keu}
\bibinfo{author}{\bibfnamefont{F.~G.} \bibnamefont{Celiberto}}
  (\bibinfo{year}{2022}{\natexlab{b}}), \eprint{2206.09413}.

\bibitem[{\citenamefont{Celiberto
  et~al.}(2022{\natexlab{e}})\citenamefont{Celiberto, Fucilla, Mohammed, and
  Papa}}]{Celiberto:2022zdg}
\bibinfo{author}{\bibfnamefont{F.~G.} \bibnamefont{Celiberto}},
  \bibinfo{author}{\bibfnamefont{M.}~\bibnamefont{Fucilla}},
  \bibinfo{author}{\bibfnamefont{M.~M.~A.} \bibnamefont{Mohammed}},
  \bibnamefont{and} \bibinfo{author}{\bibfnamefont{A.}~\bibnamefont{Papa}},
  \bibinfo{journal}{Phys. Rev. D} \textbf{\bibinfo{volume}{105}},
  \bibinfo{pages}{114056} (\bibinfo{year}{2022}{\natexlab{e}}),
  \eprint{2205.13429}.

\bibitem[{\citenamefont{Bolognino
  et~al.}(2021{\natexlab{a}})\citenamefont{Bolognino, Celiberto, Fucilla,
  Ivanov, and Papa}}]{Bolognino:2021mrc}
\bibinfo{author}{\bibfnamefont{A.~D.} \bibnamefont{Bolognino}},
  \bibinfo{author}{\bibfnamefont{F.~G.} \bibnamefont{Celiberto}},
  \bibinfo{author}{\bibfnamefont{M.}~\bibnamefont{Fucilla}},
  \bibinfo{author}{\bibfnamefont{D.~{\relax Yu}.} \bibnamefont{Ivanov}},
  \bibnamefont{and} \bibinfo{author}{\bibfnamefont{A.}~\bibnamefont{Papa}},
  \bibinfo{journal}{Phys. Rev. D} \textbf{\bibinfo{volume}{103}},
  \bibinfo{pages}{094004} (\bibinfo{year}{2021}{\natexlab{a}}),
  \eprint{2103.07396}.

\bibitem[{\citenamefont{Bolognino
  et~al.}(2022{\natexlab{c}})\citenamefont{Bolognino, Celiberto, Fucilla,
  Ivanov, and Papa}}]{Bolognino:2021hxxaux}
\bibinfo{author}{\bibfnamefont{A.~D.} \bibnamefont{Bolognino}},
  \bibinfo{author}{\bibfnamefont{F.~G.} \bibnamefont{Celiberto}},
  \bibinfo{author}{\bibfnamefont{M.}~\bibnamefont{Fucilla}},
  \bibinfo{author}{\bibfnamefont{D.~{\relax Yu}.} \bibnamefont{Ivanov}},
  \bibnamefont{and} \bibinfo{author}{\bibfnamefont{A.}~\bibnamefont{Papa}},
  \bibinfo{journal}{SciPost Phys. Proc.} \textbf{\bibinfo{volume}{8}},
  \bibinfo{pages}{068} (\bibinfo{year}{2022}{\natexlab{c}}),
  \eprint{2103.07396}.

\bibitem[{\citenamefont{Brodsky
  et~al.}(1997{\natexlab{a}})\citenamefont{Brodsky, Hautmann, and
  Soper}}]{Brodsky:1996sg}
\bibinfo{author}{\bibfnamefont{S.~J.} \bibnamefont{Brodsky}},
  \bibinfo{author}{\bibfnamefont{F.}~\bibnamefont{Hautmann}}, \bibnamefont{and}
  \bibinfo{author}{\bibfnamefont{D.~E.} \bibnamefont{Soper}},
  \bibinfo{journal}{Phys. Rev. Lett.} \textbf{\bibinfo{volume}{78}},
  \bibinfo{pages}{803} (\bibinfo{year}{1997}{\natexlab{a}}),
  \bibinfo{note}{[Erratum: Phys.Rev.Lett. 79, 3544 (1997)]},
  \eprint{hep-ph/9610260}.

\bibitem[{\citenamefont{Brodsky
  et~al.}(1997{\natexlab{b}})\citenamefont{Brodsky, Hautmann, and
  Soper}}]{Brodsky:1997sd}
\bibinfo{author}{\bibfnamefont{S.~J.} \bibnamefont{Brodsky}},
  \bibinfo{author}{\bibfnamefont{F.}~\bibnamefont{Hautmann}}, \bibnamefont{and}
  \bibinfo{author}{\bibfnamefont{D.~E.} \bibnamefont{Soper}},
  \bibinfo{journal}{Phys. Rev. D} \textbf{\bibinfo{volume}{56}},
  \bibinfo{pages}{6957} (\bibinfo{year}{1997}{\natexlab{b}}),
  \eprint{hep-ph/9706427}.

\bibitem[{\citenamefont{Brodsky et~al.}(1999)\citenamefont{Brodsky, Fadin, Kim,
  Lipatov, and Pivovarov}}]{Brodsky:1998kn}
\bibinfo{author}{\bibfnamefont{S.~J.} \bibnamefont{Brodsky}},
  \bibinfo{author}{\bibfnamefont{V.~S.} \bibnamefont{Fadin}},
  \bibinfo{author}{\bibfnamefont{V.~T.} \bibnamefont{Kim}},
  \bibinfo{author}{\bibfnamefont{L.~N.} \bibnamefont{Lipatov}},
  \bibnamefont{and} \bibinfo{author}{\bibfnamefont{G.~B.}
  \bibnamefont{Pivovarov}}, \bibinfo{journal}{JETP Lett.}
  \textbf{\bibinfo{volume}{70}}, \bibinfo{pages}{155} (\bibinfo{year}{1999}),
  \eprint{hep-ph/9901229}.

\bibitem[{\citenamefont{Brodsky et~al.}(2002)\citenamefont{Brodsky, Fadin, Kim,
  Lipatov, and Pivovarov}}]{Brodsky:2002ka}
\bibinfo{author}{\bibfnamefont{S.~J.} \bibnamefont{Brodsky}},
  \bibinfo{author}{\bibfnamefont{V.~S.} \bibnamefont{Fadin}},
  \bibinfo{author}{\bibfnamefont{V.~T.} \bibnamefont{Kim}},
  \bibinfo{author}{\bibfnamefont{L.~N.} \bibnamefont{Lipatov}},
  \bibnamefont{and} \bibinfo{author}{\bibfnamefont{G.~B.}
  \bibnamefont{Pivovarov}}, \bibinfo{journal}{JETP Lett.}
  \textbf{\bibinfo{volume}{76}}, \bibinfo{pages}{249} (\bibinfo{year}{2002}),
  \eprint{hep-ph/0207297}.

\bibitem[{\citenamefont{Celiberto}(2021{\natexlab{a}})}]{Celiberto:2020wpk}
\bibinfo{author}{\bibfnamefont{F.~G.} \bibnamefont{Celiberto}},
  \bibinfo{journal}{Eur. Phys. J. C} \textbf{\bibinfo{volume}{81}},
  \bibinfo{pages}{691} (\bibinfo{year}{2021}{\natexlab{a}}),
  \eprint{2008.07378}.

\bibitem[{\citenamefont{Hentschinski et~al.}(2021)\citenamefont{Hentschinski,
  Kutak, and van Hameren}}]{Hentschinski:2020tbi}
\bibinfo{author}{\bibfnamefont{M.}~\bibnamefont{Hentschinski}},
  \bibinfo{author}{\bibfnamefont{K.}~\bibnamefont{Kutak}}, \bibnamefont{and}
  \bibinfo{author}{\bibfnamefont{A.}~\bibnamefont{van Hameren}},
  \bibinfo{journal}{Eur. Phys. J. C} \textbf{\bibinfo{volume}{81}},
  \bibinfo{pages}{112} (\bibinfo{year}{2021}), \bibinfo{note}{[Erratum: Eur.
  Phys. J. C 81, 262 (2021)]}, \eprint{2011.03193}.

\bibitem[{\citenamefont{Mele and Nason}(1991)}]{Mele:1990cw}
\bibinfo{author}{\bibfnamefont{B.}~\bibnamefont{Mele}} \bibnamefont{and}
  \bibinfo{author}{\bibfnamefont{P.}~\bibnamefont{Nason}},
  \bibinfo{journal}{Nucl. Phys. B} \textbf{\bibinfo{volume}{361}},
  \bibinfo{pages}{626} (\bibinfo{year}{1991}), \bibinfo{note}{[Erratum:
  Nucl.Phys.B 921, 841--842 (2017)]}.

\bibitem[{\citenamefont{Cacciari and Greco}(1994)}]{Cacciari:1993mq}
\bibinfo{author}{\bibfnamefont{M.}~\bibnamefont{Cacciari}} \bibnamefont{and}
  \bibinfo{author}{\bibfnamefont{M.}~\bibnamefont{Greco}},
  \bibinfo{journal}{Nucl. Phys. B} \textbf{\bibinfo{volume}{421}},
  \bibinfo{pages}{530} (\bibinfo{year}{1994}), \eprint{hep-ph/9311260}.

\bibitem[{\citenamefont{Zheng et~al.}(2019)\citenamefont{Zheng, Chang, and
  Wu}}]{Zheng:2019dfk}
\bibinfo{author}{\bibfnamefont{X.-C.} \bibnamefont{Zheng}},
  \bibinfo{author}{\bibfnamefont{C.-H.} \bibnamefont{Chang}}, \bibnamefont{and}
  \bibinfo{author}{\bibfnamefont{X.-G.} \bibnamefont{Wu}},
  \bibinfo{journal}{Phys. Rev. D} \textbf{\bibinfo{volume}{100}},
  \bibinfo{pages}{014005} (\bibinfo{year}{2019}), \eprint{1905.09171}.

\bibitem[{\citenamefont{Boussarie et~al.}(2015)\citenamefont{Boussarie,
  Duclou\'e, Szymanowski, and Wallon}}]{Boussarie:2015jar}
\bibinfo{author}{\bibfnamefont{R.}~\bibnamefont{Boussarie}},
  \bibinfo{author}{\bibfnamefont{B.}~\bibnamefont{Duclou\'e}},
  \bibinfo{author}{\bibfnamefont{L.}~\bibnamefont{Szymanowski}},
  \bibnamefont{and} \bibinfo{author}{\bibfnamefont{S.}~\bibnamefont{Wallon}},
  in \emph{\bibinfo{booktitle}{{International Conference on the Structure and
  Interactions of the Photon and 21st International Workshop on Photon-Photon
  Collisions and International Workshop on High Energy Photon Linear
  Colliders}}} (\bibinfo{year}{2015}), \eprint{1511.02181}.

\bibitem[{\citenamefont{Boussarie et~al.}(2016)\citenamefont{Boussarie,
  Ducloue, Szymanowski, and Wallon}}]{Boussarie:2016gaq}
\bibinfo{author}{\bibfnamefont{R.}~\bibnamefont{Boussarie}},
  \bibinfo{author}{\bibfnamefont{B.}~\bibnamefont{Ducloue}},
  \bibinfo{author}{\bibfnamefont{L.}~\bibnamefont{Szymanowski}},
  \bibnamefont{and} \bibinfo{author}{\bibfnamefont{S.}~\bibnamefont{Wallon}},
  \bibinfo{journal}{PoS} \textbf{\bibinfo{volume}{DIS2016}},
  \bibinfo{pages}{204} (\bibinfo{year}{2016}).

\bibitem[{\citenamefont{Kotikov and Lipatov}(2000)}]{Kotikov:2000pm}
\bibinfo{author}{\bibfnamefont{A.~V.} \bibnamefont{Kotikov}} \bibnamefont{and}
  \bibinfo{author}{\bibfnamefont{L.~N.} \bibnamefont{Lipatov}},
  \bibinfo{journal}{Nucl. Phys. B} \textbf{\bibinfo{volume}{582}},
  \bibinfo{pages}{19} (\bibinfo{year}{2000}), \eprint{hep-ph/0004008}.

\bibitem[{\citenamefont{Kotikov and Lipatov}(2003)}]{Kotikov:2002ab}
\bibinfo{author}{\bibfnamefont{A.~V.} \bibnamefont{Kotikov}} \bibnamefont{and}
  \bibinfo{author}{\bibfnamefont{L.~N.} \bibnamefont{Lipatov}},
  \bibinfo{journal}{Nucl. Phys. B} \textbf{\bibinfo{volume}{661}},
  \bibinfo{pages}{19} (\bibinfo{year}{2003}), \bibinfo{note}{[Erratum:
  Nucl.Phys.B 685, 405--407 (2004)]}, \eprint{hep-ph/0208220}.

\bibitem[{\citenamefont{Ivanov and Papa}(2012)}]{Ivanov:2012iv}
\bibinfo{author}{\bibfnamefont{D.~{\relax Yu}.} \bibnamefont{Ivanov}}
  \bibnamefont{and} \bibinfo{author}{\bibfnamefont{A.}~\bibnamefont{Papa}},
  \bibinfo{journal}{JHEP} \textbf{\bibinfo{volume}{07}}, \bibinfo{pages}{045}
  (\bibinfo{year}{2012}), \eprint{1205.6068}.

\bibitem[{\citenamefont{Thacker and Lepage}(1991)}]{Thacker:1990bm}
\bibinfo{author}{\bibfnamefont{B.~A.} \bibnamefont{Thacker}} \bibnamefont{and}
  \bibinfo{author}{\bibfnamefont{G.~P.} \bibnamefont{Lepage}},
  \bibinfo{journal}{Phys. Rev. D} \textbf{\bibinfo{volume}{43}},
  \bibinfo{pages}{196} (\bibinfo{year}{1991}).

\bibitem[{\citenamefont{Bodwin et~al.}(1995)\citenamefont{Bodwin, Braaten, and
  Lepage}}]{Bodwin:1994jh}
\bibinfo{author}{\bibfnamefont{G.~T.} \bibnamefont{Bodwin}},
  \bibinfo{author}{\bibfnamefont{E.}~\bibnamefont{Braaten}}, \bibnamefont{and}
  \bibinfo{author}{\bibfnamefont{G.~P.} \bibnamefont{Lepage}},
  \bibinfo{journal}{Phys. Rev. D} \textbf{\bibinfo{volume}{51}},
  \bibinfo{pages}{1125} (\bibinfo{year}{1995}), \bibinfo{note}{[Erratum:
  Phys.Rev.D 55, 5853 (1997)]}, \eprint{hep-ph/9407339}.

\bibitem[{\citenamefont{Brambilla
  et~al.}(2004)}]{QuarkoniumWorkingGroup:2004kpm}
\bibinfo{author}{\bibfnamefont{N.}~\bibnamefont{Brambilla}}
  \bibnamefont{et~al.} (\bibinfo{collaboration}{Quarkonium Working Group})
  (\bibinfo{year}{2004}), \eprint{hep-ph/0412158}.

\bibitem[{\citenamefont{Brambilla et~al.}(2011)}]{Brambilla:2010cs}
\bibinfo{author}{\bibfnamefont{N.}~\bibnamefont{Brambilla}}
  \bibnamefont{et~al.}, \bibinfo{journal}{Eur. Phys. J. C}
  \textbf{\bibinfo{volume}{71}}, \bibinfo{pages}{1534} (\bibinfo{year}{2011}),
  \eprint{1010.5827}.

\bibitem[{\citenamefont{Pineda}(2012)}]{Pineda:2011dg}
\bibinfo{author}{\bibfnamefont{A.}~\bibnamefont{Pineda}},
  \bibinfo{journal}{Prog. Part. Nucl. Phys.} \textbf{\bibinfo{volume}{67}},
  \bibinfo{pages}{735} (\bibinfo{year}{2012}), \eprint{1111.0165}.

\bibitem[{\citenamefont{Brambilla et~al.}(2021)\citenamefont{Brambilla, Chung,
  and Vairo}}]{Brambilla:2020ojz}
\bibinfo{author}{\bibfnamefont{N.}~\bibnamefont{Brambilla}},
  \bibinfo{author}{\bibfnamefont{H.~S.} \bibnamefont{Chung}}, \bibnamefont{and}
  \bibinfo{author}{\bibfnamefont{A.}~\bibnamefont{Vairo}},
  \bibinfo{journal}{Phys. Rev. Lett.} \textbf{\bibinfo{volume}{126}},
  \bibinfo{pages}{082003} (\bibinfo{year}{2021}), \eprint{2007.07613}.

\bibitem[{\citenamefont{Eichten and Quigg}(1994)}]{Eichten:1994gt}
\bibinfo{author}{\bibfnamefont{E.~J.} \bibnamefont{Eichten}} \bibnamefont{and}
  \bibinfo{author}{\bibfnamefont{C.}~\bibnamefont{Quigg}},
  \bibinfo{journal}{Phys. Rev. D} \textbf{\bibinfo{volume}{49}},
  \bibinfo{pages}{5845} (\bibinfo{year}{1994}), \eprint{hep-ph/9402210}.

\bibitem[{\citenamefont{Braaten et~al.}(1993)\citenamefont{Braaten, Cheung, and
  Yuan}}]{Braaten:1993mp}
\bibinfo{author}{\bibfnamefont{E.}~\bibnamefont{Braaten}},
  \bibinfo{author}{\bibfnamefont{K.-m.} \bibnamefont{Cheung}},
  \bibnamefont{and} \bibinfo{author}{\bibfnamefont{T.~C.} \bibnamefont{Yuan}},
  \bibinfo{journal}{Phys. Rev. D} \textbf{\bibinfo{volume}{48}},
  \bibinfo{pages}{4230} (\bibinfo{year}{1993}), \eprint{hep-ph/9302307}.

\bibitem[{\citenamefont{Khachatryan et~al.}(2016)}]{Khachatryan:2016udy}
\bibinfo{author}{\bibfnamefont{V.}~\bibnamefont{Khachatryan}}
  \bibnamefont{et~al.} (\bibinfo{collaboration}{CMS}), \bibinfo{journal}{JHEP}
  \textbf{\bibinfo{volume}{08}}, \bibinfo{pages}{139} (\bibinfo{year}{2016}),
  \eprint{1601.06713}.

\bibitem[{\citenamefont{Braaten and Yuan}(1993)}]{Braaten:1993rw}
\bibinfo{author}{\bibfnamefont{E.}~\bibnamefont{Braaten}} \bibnamefont{and}
  \bibinfo{author}{\bibfnamefont{T.~C.} \bibnamefont{Yuan}},
  \bibinfo{journal}{Phys. Rev. Lett.} \textbf{\bibinfo{volume}{71}},
  \bibinfo{pages}{1673} (\bibinfo{year}{1993}), \eprint{hep-ph/9303205}.

\bibitem[{\citenamefont{Bautista et~al.}(2016)\citenamefont{Bautista,
  Fernandez~Tellez, and Hentschinski}}]{Bautista:2016xnp}
\bibinfo{author}{\bibfnamefont{I.}~\bibnamefont{Bautista}},
  \bibinfo{author}{\bibfnamefont{A.}~\bibnamefont{Fernandez~Tellez}},
  \bibnamefont{and}
  \bibinfo{author}{\bibfnamefont{M.}~\bibnamefont{Hentschinski}},
  \bibinfo{journal}{Phys. Rev. D} \textbf{\bibinfo{volume}{94}},
  \bibinfo{pages}{054002} (\bibinfo{year}{2016}), \eprint{1607.05203}.

\bibitem[{\citenamefont{Arroyo~Garcia et~al.}(2019)\citenamefont{Arroyo~Garcia,
  Hentschinski, and Kutak}}]{Garcia:2019tne}
\bibinfo{author}{\bibfnamefont{A.}~\bibnamefont{Arroyo~Garcia}},
  \bibinfo{author}{\bibfnamefont{M.}~\bibnamefont{Hentschinski}},
  \bibnamefont{and} \bibinfo{author}{\bibfnamefont{K.}~\bibnamefont{Kutak}},
  \bibinfo{journal}{Phys. Lett. B} \textbf{\bibinfo{volume}{795}},
  \bibinfo{pages}{569} (\bibinfo{year}{2019}), \eprint{1904.04394}.

\bibitem[{\citenamefont{Hentschinski and
  Padr\'on~Molina}(2021)}]{Hentschinski:2020yfm}
\bibinfo{author}{\bibfnamefont{M.}~\bibnamefont{Hentschinski}}
  \bibnamefont{and}
  \bibinfo{author}{\bibfnamefont{E.}~\bibnamefont{Padr\'on~Molina}},
  \bibinfo{journal}{Phys. Rev. D} \textbf{\bibinfo{volume}{103}},
  \bibinfo{pages}{074008} (\bibinfo{year}{2021}), \eprint{2011.02640}.

\bibitem[{\citenamefont{Hentschinski et~al.}(2013)\citenamefont{Hentschinski,
  Sabio~Vera, and Salas}}]{Hentschinski:2012kr}
\bibinfo{author}{\bibfnamefont{M.}~\bibnamefont{Hentschinski}},
  \bibinfo{author}{\bibfnamefont{A.}~\bibnamefont{Sabio~Vera}},
  \bibnamefont{and} \bibinfo{author}{\bibfnamefont{C.}~\bibnamefont{Salas}},
  \bibinfo{journal}{Phys. Rev. Lett.} \textbf{\bibinfo{volume}{110}},
  \bibinfo{pages}{041601} (\bibinfo{year}{2013}), \eprint{1209.1353}.

\bibitem[{\citenamefont{Anikin et~al.}(2010)\citenamefont{Anikin, Ivanov, Pire,
  Szymanowski, and Wallon}}]{Anikin:2009bf}
\bibinfo{author}{\bibfnamefont{I.}~\bibnamefont{Anikin}},
  \bibinfo{author}{\bibfnamefont{D.~{\relax Yu}.} \bibnamefont{Ivanov}},
  \bibinfo{author}{\bibfnamefont{B.}~\bibnamefont{Pire}},
  \bibinfo{author}{\bibfnamefont{L.}~\bibnamefont{Szymanowski}},
  \bibnamefont{and} \bibinfo{author}{\bibfnamefont{S.}~\bibnamefont{Wallon}},
  \bibinfo{journal}{Nucl. Phys. B} \textbf{\bibinfo{volume}{828}},
  \bibinfo{pages}{1} (\bibinfo{year}{2010}), \eprint{0909.4090}.

\bibitem[{\citenamefont{Anikin et~al.}(2011)\citenamefont{Anikin, Besse,
  Ivanov, Pire, Szymanowski, and Wallon}}]{Anikin:2011sa}
\bibinfo{author}{\bibfnamefont{I.}~\bibnamefont{Anikin}},
  \bibinfo{author}{\bibfnamefont{A.}~\bibnamefont{Besse}},
  \bibinfo{author}{\bibfnamefont{D.~{\relax Yu}.} \bibnamefont{Ivanov}},
  \bibinfo{author}{\bibfnamefont{B.}~\bibnamefont{Pire}},
  \bibinfo{author}{\bibfnamefont{L.}~\bibnamefont{Szymanowski}},
  \bibnamefont{and} \bibinfo{author}{\bibfnamefont{S.}~\bibnamefont{Wallon}},
  \bibinfo{journal}{Phys. Rev. D} \textbf{\bibinfo{volume}{84}},
  \bibinfo{pages}{054004} (\bibinfo{year}{2011}), \eprint{1105.1761}.

\bibitem[{\citenamefont{Besse et~al.}(2013)\citenamefont{Besse, Szymanowski,
  and Wallon}}]{Besse:2013muy}
\bibinfo{author}{\bibfnamefont{A.}~\bibnamefont{Besse}},
  \bibinfo{author}{\bibfnamefont{L.}~\bibnamefont{Szymanowski}},
  \bibnamefont{and} \bibinfo{author}{\bibfnamefont{S.}~\bibnamefont{Wallon}},
  \bibinfo{journal}{JHEP} \textbf{\bibinfo{volume}{11}}, \bibinfo{pages}{062}
  (\bibinfo{year}{2013}), \eprint{1302.1766}.

\bibitem[{\citenamefont{Bolognino
  et~al.}(2018{\natexlab{b}})\citenamefont{Bolognino, Celiberto, Ivanov, and
  Papa}}]{Bolognino:2018rhb}
\bibinfo{author}{\bibfnamefont{A.~D.} \bibnamefont{Bolognino}},
  \bibinfo{author}{\bibfnamefont{F.~G.} \bibnamefont{Celiberto}},
  \bibinfo{author}{\bibfnamefont{D.~{\relax Yu}.} \bibnamefont{Ivanov}},
  \bibnamefont{and} \bibinfo{author}{\bibfnamefont{A.}~\bibnamefont{Papa}},
  \bibinfo{journal}{Eur. Phys. J.} \textbf{\bibinfo{volume}{C78}},
  \bibinfo{pages}{1023} (\bibinfo{year}{2018}{\natexlab{b}}),
  \eprint{1808.02395}.

\bibitem[{\citenamefont{Bolognino
  et~al.}(2018{\natexlab{c}})\citenamefont{Bolognino, Celiberto, Ivanov, and
  Papa}}]{Bolognino:2018mlw}
\bibinfo{author}{\bibfnamefont{A.~D.} \bibnamefont{Bolognino}},
  \bibinfo{author}{\bibfnamefont{F.~G.} \bibnamefont{Celiberto}},
  \bibinfo{author}{\bibfnamefont{D.~{\relax Yu}.} \bibnamefont{Ivanov}},
  \bibnamefont{and} \bibinfo{author}{\bibfnamefont{A.}~\bibnamefont{Papa}},
  \bibinfo{journal}{Frascati Phys. Ser.} \textbf{\bibinfo{volume}{67}},
  \bibinfo{pages}{76} (\bibinfo{year}{2018}{\natexlab{c}}),
  \eprint{1808.02958}.

\bibitem[{\citenamefont{Bolognino
  et~al.}(2019{\natexlab{f}})\citenamefont{Bolognino, Celiberto, Ivanov, and
  Papa}}]{Bolognino:2019bko}
\bibinfo{author}{\bibfnamefont{A.~D.} \bibnamefont{Bolognino}},
  \bibinfo{author}{\bibfnamefont{F.~G.} \bibnamefont{Celiberto}},
  \bibinfo{author}{\bibfnamefont{D.~{\relax Yu}.} \bibnamefont{Ivanov}},
  \bibnamefont{and} \bibinfo{author}{\bibfnamefont{A.}~\bibnamefont{Papa}},
  \bibinfo{journal}{Acta Phys. Polon. Supp.} \textbf{\bibinfo{volume}{12}},
  \bibinfo{pages}{891} (\bibinfo{year}{2019}{\natexlab{f}}),
  \eprint{1902.04520}.

\bibitem[{\citenamefont{Bolognino et~al.}(2020)\citenamefont{Bolognino,
  Szczurek, and Schaefer}}]{Bolognino:2019pba}
\bibinfo{author}{\bibfnamefont{A.~D.} \bibnamefont{Bolognino}},
  \bibinfo{author}{\bibfnamefont{A.}~\bibnamefont{Szczurek}}, \bibnamefont{and}
  \bibinfo{author}{\bibfnamefont{W.}~\bibnamefont{Schaefer}},
  \bibinfo{journal}{Phys. Rev. D} \textbf{\bibinfo{volume}{101}},
  \bibinfo{pages}{054041} (\bibinfo{year}{2020}), \eprint{1912.06507}.

\bibitem[{\citenamefont{Celiberto}(2019)}]{Celiberto:2019slj}
\bibinfo{author}{\bibfnamefont{F.~G.} \bibnamefont{Celiberto}},
  \bibinfo{journal}{Nuovo Cim.} \textbf{\bibinfo{volume}{C42}},
  \bibinfo{pages}{220} (\bibinfo{year}{2019}), \eprint{1912.11313}.

\bibitem[{\citenamefont{Bolognino
  et~al.}(2021{\natexlab{b}})\citenamefont{Bolognino, Celiberto, Ivanov, Papa,
  Sch\"afer, and Szczurek}}]{Bolognino:2021niq}
\bibinfo{author}{\bibfnamefont{A.~D.} \bibnamefont{Bolognino}},
  \bibinfo{author}{\bibfnamefont{F.~G.} \bibnamefont{Celiberto}},
  \bibinfo{author}{\bibfnamefont{D.~{\relax Yu}.} \bibnamefont{Ivanov}},
  \bibinfo{author}{\bibfnamefont{A.}~\bibnamefont{Papa}},
  \bibinfo{author}{\bibfnamefont{W.}~\bibnamefont{Sch\"afer}},
  \bibnamefont{and} \bibinfo{author}{\bibfnamefont{A.}~\bibnamefont{Szczurek}},
  \bibinfo{journal}{Eur. Phys. J. C} \textbf{\bibinfo{volume}{81}},
  \bibinfo{pages}{846} (\bibinfo{year}{2021}{\natexlab{b}}),
  \eprint{2107.13415}.

\bibitem[{\citenamefont{Bolognino
  et~al.}(2022{\natexlab{d}})\citenamefont{Bolognino, Celiberto, Ivanov, and
  Papa}}]{Bolognino:2021gjm}
\bibinfo{author}{\bibfnamefont{A.~D.} \bibnamefont{Bolognino}},
  \bibinfo{author}{\bibfnamefont{F.~G.} \bibnamefont{Celiberto}},
  \bibinfo{author}{\bibfnamefont{D.~Y.} \bibnamefont{Ivanov}},
  \bibnamefont{and} \bibinfo{author}{\bibfnamefont{A.}~\bibnamefont{Papa}},
  \bibinfo{journal}{SciPost Phys. Proc.} \textbf{\bibinfo{volume}{8}},
  \bibinfo{pages}{089} (\bibinfo{year}{2022}{\natexlab{d}}),
  \eprint{2107.12725}.

\bibitem[{\citenamefont{Bolognino
  et~al.}(2022{\natexlab{e}})\citenamefont{Bolognino, Celiberto, Fucilla,
  Ivanov, Papa, Sch\"afer, and Szczurek}}]{Bolognino:2022uty}
\bibinfo{author}{\bibfnamefont{A.~D.} \bibnamefont{Bolognino}},
  \bibinfo{author}{\bibfnamefont{F.~G.} \bibnamefont{Celiberto}},
  \bibinfo{author}{\bibfnamefont{M.}~\bibnamefont{Fucilla}},
  \bibinfo{author}{\bibfnamefont{D.~{\relax Yu}.} \bibnamefont{Ivanov}},
  \bibinfo{author}{\bibfnamefont{A.}~\bibnamefont{Papa}},
  \bibinfo{author}{\bibfnamefont{W.}~\bibnamefont{Sch\"afer}},
  \bibnamefont{and} \bibinfo{author}{\bibfnamefont{A.}~\bibnamefont{Szczurek}},
  in \emph{\bibinfo{booktitle}{{19th International Conference on Hadron
  Spectroscopy and Structure}}} (\bibinfo{year}{2022}{\natexlab{e}}),
  \eprint{2202.02513}.

\bibitem[{\citenamefont{Celiberto}(2022{\natexlab{c}})}]{Celiberto:2022fam}
\bibinfo{author}{\bibfnamefont{F.~G.} \bibnamefont{Celiberto}}
  (\bibinfo{year}{2022}{\natexlab{c}}), \eprint{2202.04207}.

\bibitem[{\citenamefont{Bolognino
  et~al.}(2022{\natexlab{f}})\citenamefont{Bolognino, Celiberto, Ivanov, Papa,
  Sch\"afer, and Szczurek}}]{Bolognino:2022ndh}
\bibinfo{author}{\bibfnamefont{A.~D.} \bibnamefont{Bolognino}},
  \bibinfo{author}{\bibfnamefont{F.~G.} \bibnamefont{Celiberto}},
  \bibinfo{author}{\bibfnamefont{D.~{\relax Yu}.} \bibnamefont{Ivanov}},
  \bibinfo{author}{\bibfnamefont{A.}~\bibnamefont{Papa}},
  \bibinfo{author}{\bibfnamefont{W.}~\bibnamefont{Sch\"afer}},
  \bibnamefont{and} \bibinfo{author}{\bibfnamefont{A.}~\bibnamefont{Szczurek}},
  \bibinfo{journal}{Zenodo, in press}  (\bibinfo{year}{2022}{\natexlab{f}}),
  \eprint{2207.05726}.

\bibitem[{\citenamefont{Motyka et~al.}(2015)\citenamefont{Motyka, Sadzikowski,
  and Stebel}}]{Motyka:2014lya}
\bibinfo{author}{\bibfnamefont{L.}~\bibnamefont{Motyka}},
  \bibinfo{author}{\bibfnamefont{M.}~\bibnamefont{Sadzikowski}},
  \bibnamefont{and} \bibinfo{author}{\bibfnamefont{T.}~\bibnamefont{Stebel}},
  \bibinfo{journal}{JHEP} \textbf{\bibinfo{volume}{05}}, \bibinfo{pages}{087}
  (\bibinfo{year}{2015}), \eprint{1412.4675}.

\bibitem[{\citenamefont{Brzeminski et~al.}(2017)\citenamefont{Brzeminski,
  Motyka, Sadzikowski, and Stebel}}]{Brzeminski:2016lwh}
\bibinfo{author}{\bibfnamefont{D.}~\bibnamefont{Brzeminski}},
  \bibinfo{author}{\bibfnamefont{L.}~\bibnamefont{Motyka}},
  \bibinfo{author}{\bibfnamefont{M.}~\bibnamefont{Sadzikowski}},
  \bibnamefont{and} \bibinfo{author}{\bibfnamefont{T.}~\bibnamefont{Stebel}},
  \bibinfo{journal}{JHEP} \textbf{\bibinfo{volume}{01}}, \bibinfo{pages}{005}
  (\bibinfo{year}{2017}), \eprint{1611.04449}.

\bibitem[{\citenamefont{Motyka et~al.}(2017)\citenamefont{Motyka, Sadzikowski,
  and Stebel}}]{Motyka:2016lta}
\bibinfo{author}{\bibfnamefont{L.}~\bibnamefont{Motyka}},
  \bibinfo{author}{\bibfnamefont{M.}~\bibnamefont{Sadzikowski}},
  \bibnamefont{and} \bibinfo{author}{\bibfnamefont{T.}~\bibnamefont{Stebel}},
  \bibinfo{journal}{Phys. Rev.} \textbf{\bibinfo{volume}{D95}},
  \bibinfo{pages}{114025} (\bibinfo{year}{2017}), \eprint{1609.04300}.

\bibitem[{\citenamefont{Celiberto
  et~al.}(2018{\natexlab{b}})\citenamefont{Celiberto, Gordo~G\'omez, and
  Sabio~Vera}}]{Celiberto:2018muu}
\bibinfo{author}{\bibfnamefont{F.~G.} \bibnamefont{Celiberto}},
  \bibinfo{author}{\bibfnamefont{D.}~\bibnamefont{Gordo~G\'omez}},
  \bibnamefont{and}
  \bibinfo{author}{\bibfnamefont{A.}~\bibnamefont{Sabio~Vera}},
  \bibinfo{journal}{Phys. Lett.} \textbf{\bibinfo{volume}{B786}},
  \bibinfo{pages}{201} (\bibinfo{year}{2018}{\natexlab{b}}),
  \eprint{1808.09511}.

\bibitem[{\citenamefont{Bacchetta et~al.}(2020)\citenamefont{Bacchetta,
  Celiberto, Radici, and Taels}}]{Bacchetta:2020vty}
\bibinfo{author}{\bibfnamefont{A.}~\bibnamefont{Bacchetta}},
  \bibinfo{author}{\bibfnamefont{F.~G.} \bibnamefont{Celiberto}},
  \bibinfo{author}{\bibfnamefont{M.}~\bibnamefont{Radici}}, \bibnamefont{and}
  \bibinfo{author}{\bibfnamefont{P.}~\bibnamefont{Taels}},
  \bibinfo{journal}{Eur. Phys. J. C} \textbf{\bibinfo{volume}{80}},
  \bibinfo{pages}{733} (\bibinfo{year}{2020}), \eprint{2005.02288}.

\bibitem[{\citenamefont{Celiberto}(2021{\natexlab{b}})}]{Celiberto:2021zww}
\bibinfo{author}{\bibfnamefont{F.~G.} \bibnamefont{Celiberto}},
  \bibinfo{journal}{Nuovo Cim.} \textbf{\bibinfo{volume}{C44}},
  \bibinfo{pages}{36} (\bibinfo{year}{2021}{\natexlab{b}}),
  \eprint{2101.04630}.

\bibitem[{\citenamefont{Bacchetta
  et~al.}(2022{\natexlab{a}})\citenamefont{Bacchetta, Celiberto, Radici, and
  Taels}}]{Bacchetta:2021oht}
\bibinfo{author}{\bibfnamefont{A.}~\bibnamefont{Bacchetta}},
  \bibinfo{author}{\bibfnamefont{F.~G.} \bibnamefont{Celiberto}},
  \bibinfo{author}{\bibfnamefont{M.}~\bibnamefont{Radici}}, \bibnamefont{and}
  \bibinfo{author}{\bibfnamefont{P.}~\bibnamefont{Taels}},
  \bibinfo{journal}{SciPost Phys. Proc.} \textbf{\bibinfo{volume}{8}},
  \bibinfo{pages}{040} (\bibinfo{year}{2022}{\natexlab{a}}),
  \eprint{2107.13446}.

\bibitem[{\citenamefont{Bacchetta
  et~al.}(2022{\natexlab{b}})\citenamefont{Bacchetta, Celiberto, and
  Radici}}]{Bacchetta:2021lvw}
\bibinfo{author}{\bibfnamefont{A.}~\bibnamefont{Bacchetta}},
  \bibinfo{author}{\bibfnamefont{F.~G.} \bibnamefont{Celiberto}},
  \bibnamefont{and} \bibinfo{author}{\bibfnamefont{M.}~\bibnamefont{Radici}},
  \bibinfo{journal}{PoS} \textbf{\bibinfo{volume}{EPS-HEP2021}},
  \bibinfo{pages}{376} (\bibinfo{year}{2022}{\natexlab{b}}),
  \eprint{2111.01686}.

\bibitem[{\citenamefont{Bacchetta
  et~al.}(2022{\natexlab{c}})\citenamefont{Bacchetta, Celiberto, and
  Radici}}]{Bacchetta:2021twk}
\bibinfo{author}{\bibfnamefont{A.}~\bibnamefont{Bacchetta}},
  \bibinfo{author}{\bibfnamefont{F.~G.} \bibnamefont{Celiberto}},
  \bibnamefont{and} \bibinfo{author}{\bibfnamefont{M.}~\bibnamefont{Radici}},
  \bibinfo{journal}{PoS} \textbf{\bibinfo{volume}{PANIC2021}},
  \bibinfo{pages}{378} (\bibinfo{year}{2022}{\natexlab{c}}),
  \eprint{2111.03567}.

\bibitem[{\citenamefont{Bacchetta
  et~al.}(2022{\natexlab{d}})\citenamefont{Bacchetta, Celiberto, and
  Radici}}]{Bacchetta:2022esb}
\bibinfo{author}{\bibfnamefont{A.}~\bibnamefont{Bacchetta}},
  \bibinfo{author}{\bibfnamefont{F.~G.} \bibnamefont{Celiberto}},
  \bibnamefont{and} \bibinfo{author}{\bibfnamefont{M.}~\bibnamefont{Radici}}
  (\bibinfo{year}{2022}{\natexlab{d}}), \eprint{2201.10508}.

\bibitem[{\citenamefont{Bacchetta
  et~al.}(2022{\natexlab{e}})\citenamefont{Bacchetta, Celiberto, and
  Radici}}]{Bacchetta:2022crh}
\bibinfo{author}{\bibfnamefont{A.}~\bibnamefont{Bacchetta}},
  \bibinfo{author}{\bibfnamefont{F.~G.} \bibnamefont{Celiberto}},
  \bibnamefont{and} \bibinfo{author}{\bibfnamefont{M.}~\bibnamefont{Radici}}
  (\bibinfo{year}{2022}{\natexlab{e}}), \eprint{2206.07815}.

\bibitem[{\citenamefont{Bacchetta
  et~al.}(2022{\natexlab{f}})\citenamefont{Bacchetta, Celiberto, Radici, and
  Signori}}]{Bacchetta:2022nyv}
\bibinfo{author}{\bibfnamefont{A.}~\bibnamefont{Bacchetta}},
  \bibinfo{author}{\bibfnamefont{F.~G.} \bibnamefont{Celiberto}},
  \bibinfo{author}{\bibfnamefont{M.}~\bibnamefont{Radici}}, \bibnamefont{and}
  \bibinfo{author}{\bibfnamefont{A.}~\bibnamefont{Signori}}
  (\bibinfo{year}{2022}{\natexlab{f}}), \eprint{2208.06252}.

\end{thebibliography}

\end{document}